\documentclass[twocolumn,twocolappendix,acknowledgments]{AAStemp_v7/aastex7}

\usepackage{graphicx} 
\usepackage{appendix}
\usepackage{bm}	
\usepackage{booktabs}
\usepackage{threeparttable}
\usepackage{makecell}
\usepackage{multirow}
\usepackage{soul}
\usepackage{enumitem}

\usepackage{amsmath}
\usepackage{lineno}

\usepackage[dvipsnames]{xcolor}
\usepackage[utf8]{inputenc}
\usepackage{hyperref}
\usepackage{xspace}

\newcommand{\empdf}{\textsc{emPDF}\xspace}
\newcommand{\opdf}{\textsc{oPDF}\xspace}
\newcommand{\simDF}{\textsc{simDF}}

\shorttitle{data-driven measurement of MW halo mass in DESI}
\shortauthors{Zhou et al.}
 
\graphicspath{{./}{figure/}}

\begin{document}

\title{Constraining the Milky Way Mass profile with data driven distribution function using DESI DR2 halo stars}

\author{Yanrui Zhou}
\affiliation{Department of Astronomy, School of Physics and , Shanghai 200240, China\\}
\affiliation{ State Key Laboratory of Dark Matter Physics, School of Physics and Astronomy,Shanghai Jiao Tong University, Shanghai 200240, China\\}
\email{123zyr@sjtu.edu.cn}

\author{Jiaxin Han}
\thanks{jiaxin.han@sjtu.edu.cn}
\affiliation{Department of Astronomy, School of Physics and , Shanghai 200240, China\\}
\affiliation{ State Key Laboratory of Dark Matter Physics, School of Physics and Astronomy,Shanghai Jiao Tong University, Shanghai 200240, China\\}
\email{jiaxin.han@sjtu.edu.cn}

\author{Wenting Wang}
\thanks{wenting.wang@sjtu.edu.cn}
\affiliation{Department of Astronomy, School of Physics and , Shanghai 200240, China\\}
\affiliation{ State Key Laboratory of Dark Matter Physics, School of Physics and Astronomy,Shanghai Jiao Tong University, Shanghai 200240, China\\}
\email{wenting.wang@sjtu.edu.cn}

\author{Zhaozhou Li}
\thanks{zhaozhou.li@nju.edu.cn}
\affiliation{School of Astronomy and Space Science, Nanjing University, Nanjing 210093, China}
\affiliation{Key Laboratory of Modern Astronomy and Astrophysics, Nanjing University, Ministry of Education, Nanjing 210093, China}
\affiliation{Centre for Astrophysics and Planetary Science, Racah Institute of Physics, The Hebrew University, Jerusalem, 91904, Israel}
\email{zhaozhou.li@nju.edu.cn}

\author{Alexander~H.~Riley}
\affiliation{Lund Observatory, Division of Astrophysics, Department of Physics, Lund University, SE-221 00 Lund, Sweden}
\affiliation{Institute for Computational Cosmology, Department of Physics, Durham University, South Road, Durham DH1 3LE, UK}
\email{alexander.riley2@durham.ac.uk}

\author{Monica Valluri}
\affiliation{Department of Astronomy, University of Michigan, Ann Arbor, MI 48109, USA}
\affiliation{University of Michigan, 500 S. State Street, Ann Arbor, MI 48109, USA}
\email{mvalluri@umich.edu}

\author{Amanda Byström}
\affiliation{Institute for Astronomy, University of Edinburgh, Royal Observatory, Blackford Hill, Edinburgh EH9 3HJ, UK}
\email{Amanda.Bystrom@ed.ac.uk}

\author{Gustavo~E.~Medina}
\affiliation{Kapteyn Astronomical Institute, University of Groningen, P.O.Box 880, 9700 AV Groningen, The Netherlands}
\affiliation{Department of Astronomy and Astrophysics, University of Toronto, 50 St. George Street, Toronto ON, M5S 3H4, Canada}
\email{gustavo.medina@utoronto.ca}

\author{Sergey E. Koposov}
\affiliation{Institute for Astronomy, University of Edinburgh, Royal Observatory, Blackford Hill, Edinburgh EH9 3HJ, UK}
\affiliation{Institute of Astronomy, University of Cambridge, Madingley Road, Cambridge CB3 0HA, UK}
\email{Sergey.Koposov@ed.ac.uk}

\author{Ting S. Li}
\affiliation{Department of Astronomy and Astrophysics, University of Toronto, 50 St. George Street, Toronto ON, M5S 3H4, Canada}
\email{ting.li@astro.utoronto.ca}

\author[0000-0002-0740-1507]{Leandro {Beraldo e Silva}}
\affiliation{Observatório Nacional, Rio de Janeiro - RJ, 20921-400,
  Brasil}
\email{lberaldoesilva@on.br}

\author[0000-0001-9852-9954]{Oleg~Y.~Gnedin}
\affiliation{Department of Astronomy, University of Michigan, Ann Arbor, MI 48109, USA}
\affiliation{University of Michigan, 500 S. State Street, Ann Arbor, MI 48109, USA}
\email{ognedin@umich.edu}

\author[0000-0003-0853-8887]{N.~Kizhuprakkat}
\affiliation{Institute of Astronomy and Department of Physics, National Tsing Hua University, 101 Kuang-Fu Rd. Sec. 2, Hsinchu 30013, Taiwan}
\email{namitha96@gapp.nthu.edu.tw}

\author{N.~Sandford}
\affiliation{Department of Astronomy and Astrophysics, University of Toronto, 50 St. George Street, Toronto ON, M5S 3H4, Canada}
\email{nathan.sandford@utoronto.ca}

\author[0000-0002-2338-716X]{C.~S.~Frenk}
\affiliation{Institute for Computational Cosmology, Department of Physics, Durham University, South Road, Durham DH1 3LE, UK}
\email{c.s.frenk@durham.ac.uk}

\author{Binbin Gao}
\affiliation{Department of Astronomy, School of Physics and , Shanghai 200240, China\\}
\affiliation{ State Key Laboratory of Dark Matter Physics, School of Physics and Astronomy,Shanghai Jiao Tong University, Shanghai 200240, China\\}
\email{binbin_gao@sjtu.edu.cn}

\author{Yiran Wei}
\affiliation{Department of Astronomy, School of Physics and , Shanghai 200240, China\\}
\affiliation{ State Key Laboratory of Dark Matter Physics, School of Physics and Astronomy,Shanghai Jiao Tong University, Shanghai 200240, China\\}
\email{yiran.wei@sjtu.edu.cn}


\author{J.~Aguilar}
\affiliation{Lawrence Berkeley National Laboratory, 1 Cyclotron Road, Berkeley, CA 94720, USA}
\email{jaguilar\@lbl.gov}

\author[0000-0001-6098-7247]{S.~Ahlen}
\affiliation{Department of Physics, Boston University, 590 Commonwealth Avenue, Boston, MA 02215 USA}
\email{ahlen\@bu.edu}

\author[0000-0002-0084-572X]{C.~Allende~Prieto}
\affiliation{Departamento de Astrof\'{\i}sica, Universidad de La Laguna (ULL), E-38206, La Laguna, Tenerife, Spain}
\affiliation{Instituto de Astrof\'{\i}sica de Canarias, C/ V\'{\i}a L\'{a}ctea, s/n, E-38205 La Laguna, Tenerife, Spain}
\email{carlos.allende.prieto\@iac.es}

\author[0000-0003-0467-5438]{F.~Beutler}
\affiliation{Institute for Astronomy, University of Edinburgh, Royal Observatory, Blackford Hill, Edinburgh EH9 3HJ, UK}
\email{florian.beutler\@ed.ac.uk}

\author[0000-0001-9712-0006]{D.~Bianchi}
\affiliation{Dipartimento di Fisica ``Aldo Pontremoli'', Universit\`a degli Studi di Milano, Via Celoria 16, I-20133 Milano, Italy}
\affiliation{INAF-Osservatorio Astronomico di Brera, Via Brera 28, 20122 Milano, Italy}
\email{davide.bianchi1\@unimi.it}

\author{D.~Brooks}
\affiliation{Department of Physics \& Astronomy, University College London, Gower Street, London, WC1E 6BT, UK}
\email{david.brooks\@ucl.ac.uk}

\author[0000-0003-3044-5150]{A.~Carnero Rosell}
\affiliation{Departamento de Astrof\'{\i}sica, Universidad de La Laguna (ULL), E-38206, La Laguna, Tenerife, Spain}
\affiliation{Instituto de Astrof\'{\i}sica de Canarias, C/ V\'{\i}a L\'{a}ctea, s/n, E-38205 La Laguna, Tenerife, Spain}
\email{acarnero\@iac.es}

\author{T.~Claybaugh}
\affiliation{Lawrence Berkeley National Laboratory, 1 Cyclotron Road, Berkeley, CA 94720, USA}
\email{tmclaybaugh\@lbl.gov}

\author[0000-0001-8274-158X]{A.~P.~Cooper}
\affiliation{Institute of Astronomy and Department of Physics, National Tsing Hua University, 101 Kuang-Fu Rd. Sec. 2, Hsinchu 30013, Taiwan}
\email{apcooper\@gapp.nthu.edu.tw}

\author[0000-0002-1769-1640]{A.~de la Macorra}
\affiliation{Instituto de F\'{\i}sica, Universidad Nacional Aut\'{o}noma de M\'{e}xico, Circuito de la Investigaci\'{o}n Cient\'{\i}fica, Ciudad Universitaria, Cd. de M\'{e}xico C.~P.~04510, M\'{e}xico}
\email{macorra\@fisica.unam.mx}

\author[0000-0002-4928-4003]{Arjun~Dey}
\affiliation{NSF NOIRLab, 950 N. Cherry Ave., Tucson, AZ 85719, USA}
\email{arjun.dey\@noirlab.edu}

\author[0000-0002-5665-7912]{Biprateep~Dey}
\affiliation{Department of Astronomy \& Astrophysics, University of Toronto, Toronto, ON M5S 3H4, Canada}
\affiliation{Department of Physics \& Astronomy and Pittsburgh Particle Physics, Astrophysics, and Cosmology Center (PITT PACC), University of Pittsburgh, 3941 O'Hara Street, Pittsburgh, PA 15260, USA}
\email{b.dey\@utoronto.ca}

\author[0000-0002-2890-3725]{J.~E.~Forero-Romero}
\affiliation{Departamento de F\'isica, Universidad de los Andes, Cra. 1 No. 18A-10, Edificio Ip, CP 111711, Bogot\'a, Colombia}
\affiliation{Observatorio Astron\'omico, Universidad de los Andes, Cra. 1 No. 18A-10, Edificio H, CP 111711 Bogot\'a, Colombia}
\email{je.forero\@uniandes.edu.co}

\author[0000-0001-9632-0815]{E.~Gaztañaga}
\affiliation{Institut d'Estudis Espacials de Catalunya (IEEC), c/ Esteve Terradas 1, Edifici RDIT, Campus PMT-UPC, 08860 Castelldefels, Spain}
\affiliation{Institute of Cosmology and Gravitation, University of Portsmouth, Dennis Sciama Building, Portsmouth, PO1 3FX, UK}
\affiliation{Institute of Space Sciences, ICE-CSIC, Campus UAB, Carrer de Can Magrans s/n, 08913 Bellaterra, Barcelona, Spain}
\email{gaztanaga\@gmail.com}

\author[0000-0003-3142-233X]{Satya~{Gontcho A Gontcho}}
\affiliation{University of Virginia, Department of Astronomy, Charlottesville, VA 22904, USA}
\email{satya\@virginia.edu}

\author{G.~Gutierrez}
\affiliation{Fermi National Accelerator Laboratory, PO Box 500, Batavia, IL 60510, USA}
\email{gaston\@fnal.gov}

\author[0000-0002-6024-466X]{M.~Ishak}
\affiliation{Department of Physics, The University of Texas at Dallas, 800 W. Campbell Rd., Richardson, TX 75080, USA}
\email{mishak\@utdallas.edu}

\author[0000-0002-0000-2394]{S.~Juneau}
\affiliation{NSF NOIRLab, 950 N. Cherry Ave., Tucson, AZ 85719, USA}
\email{stephanie.juneau\@noirlab.edu}

\author[0000-0002-5652-8870]{T.~Karim}
\affiliation{Department of Astronomy \& Astrophysics, University of Toronto, Toronto, ON M5S 3H4, Canada}
\email{tanveer.karim\@utoronto.ca}

\author{R.~Kehoe}
\affiliation{Department of Physics, Southern Methodist University, 3215 Daniel Avenue, Dallas, TX 75275, USA}
\email{kehoe\@physics.smu.edu}

\author[0000-0002-1134-9035]{O.~Lahav}
\affiliation{Department of Physics \& Astronomy, University College London, Gower Street, London, WC1E 6BT, UK}
\email{o.lahav\@ucl.ac.uk}

\author{A.~Lambert}
\affiliation{Lawrence Berkeley National Laboratory, 1 Cyclotron Road, Berkeley, CA 94720, USA}
\email{arlambert\@lbl.gov}

\author[0000-0003-1838-8528]{M.~Landriau}
\affiliation{Lawrence Berkeley National Laboratory, 1 Cyclotron Road, Berkeley, CA 94720, USA}
\email{mlandriau\@lbl.gov}

\author[0000-0001-7178-8868]{L.~Le~Guillou}
\affiliation{Sorbonne Universit\'{e}, CNRS/IN2P3, Laboratoire de Physique Nucl\'{e}aire et de Hautes Energies (LPNHE), FR-75005 Paris, France}
\email{llg\@lpnhe.in2p3.fr}

\author[0000-0002-3677-3617]{A.~Leauthaud}
\affiliation{Department of Astronomy and Astrophysics, UCO/Lick Observatory, University of California, 1156 High Street, Santa Cruz, CA 95064, USA}
\affiliation{Department of Astronomy and Astrophysics, University of California, Santa Cruz, 1156 High Street, Santa Cruz, CA 95065, USA}
\email{alexie\@ucsc.edu}

\author[0000-0003-1887-1018]{M.~E.~Levi}
\affiliation{Lawrence Berkeley National Laboratory, 1 Cyclotron Road, Berkeley, CA 94720, USA}
\email{melevi\@lbl.gov}

\author[0000-0002-1125-7384]{A.~Meisner}
\affiliation{NSF NOIRLab, 950 N. Cherry Ave., Tucson, AZ 85719, USA}
\email{aaron.meisner\@noirlab.edu}

\author{R.~Miquel}
\affiliation{Instituci\'{o} Catalana de Recerca i Estudis Avan\c{c}ats, Passeig de Llu\'{\i}s Companys, 23, 08010 Barcelona, Spain}
\affiliation{Institut de F\'{i}sica d’Altes Energies (IFAE), The Barcelona Institute of Science and Technology, Edifici Cn, Campus UAB, 08193, Bellaterra (Barcelona), Spain}
\email{rmiquel\@ifae.es}

\author[0000-0001-9070-3102]{S.~Nadathur}
\affiliation{Institute of Cosmology and Gravitation, University of Portsmouth, Dennis Sciama Building, Portsmouth, PO1 3FX, UK}
\email{seshadri.nadathur\@port.ac.uk}

\author[0000-0002-4637-2868]{E.~Paillas}
\affiliation{Instituto de Estudios Astrof\'isicos, Facultad de Ingenier\'ia y Ciencias, Universidad Diego Portales, Av. Ej\'ercito Libertador 441, Santiago, Chile}
\affiliation{Steward Observatory, University of Arizona, 933 N. Cherry Avenue, Tucson, AZ 85721, USA}
\email{enrique.paillas\@udp.cl}
\author[0000-0003-3188-784X]{N.~Palanque-Delabrouille}
\affiliation{IRFU, CEA, Universit\'{e} Paris-Saclay, F-91191 Gif-sur-Yvette, France}
\affiliation{Lawrence Berkeley National Laboratory, 1 Cyclotron Road, Berkeley, CA 94720, USA}
\email{npalanque-delabrouille\@lbl.gov}

\author[0000-0002-0644-5727]{W.~J.~Percival}
\affiliation{Department of Physics and Astronomy, University of Waterloo, 200 University Ave W, Waterloo, ON N2L 3G1, Canada}
\affiliation{Perimeter Institute for Theoretical Physics, 31 Caroline St. North, Waterloo, ON N2L 2Y5, Canada}
\affiliation{Waterloo Centre for Astrophysics, University of Waterloo, 200 University Ave W, Waterloo, ON N2L 3G1, Canada}
\email{will.percival\@uwaterloo.ca}

\author[0000-0001-7145-8674]{F.~Prada}
\affiliation{Instituto de Astrof\'{i}sica de Andaluc\'{i}a (CSIC), Glorieta de la Astronom\'{i}a, s/n, E-18008 Granada, Spain}
\email{fprada\@iaa.es}

\author[0000-0001-6979-0125]{I.~P\'erez-R\`afols}
\affiliation{Departament de F\'isica, EEBE, Universitat Polit\`ecnica de Catalunya, c/Eduard Maristany 10, 08930 Barcelona, Spain}
\email{ignasi.perez.rafols\@upc.edu}

\author[0000-0002-3500-6635]{C.~Ravoux}
\affiliation{Universit\'{e} Clermont-Auvergne, CNRS, LPCA, 63000 Clermont-Ferrand, France}
\email{corentin.ravoux\@clermont.in2p3.fr}

\author{G.~Rossi}
\affiliation{Department of Physics and Astronomy, Sejong University, 209 Neungdong-ro, Gwangjin-gu, Seoul 05006, Republic of Korea}
\email{graziano\@sejong.ac.kr}

\author[0000-0002-9646-8198]{E.~Sanchez}
\affiliation{CIEMAT, Avenida Complutense 40, E-28040 Madrid, Spain}
\email{eusebio.sanchez\@ciemat.es}

\author[0000-0002-0408-5633]{C.~Saulder}
\affiliation{Max Planck Institute for Extraterrestrial Physics, Gie\ss enbachstra\ss e 1, 85748 Garching, Germany}
\email{csaulder\@mpe.mpg.de}

\author{D.~Schlegel}
\affiliation{Lawrence Berkeley National Laboratory, 1 Cyclotron Road, Berkeley, CA 94720, USA}
\email{djschlegel\@lbl.gov}

\author[0000-0002-3461-0320]{J.~Silber}
\affiliation{Lawrence Berkeley National Laboratory, 1 Cyclotron Road, Berkeley, CA 94720, USA}
\email{jhsilber\@lbl.gov}

\author[0000-0002-2949-2155]{M.~Siudek}
\affiliation{Institute of Space Sciences, ICE-CSIC, Campus UAB, Carrer de Can Magrans s/n, 08913 Bellaterra, Barcelona, Spain}
\affiliation{Instituto de Astrof\'{\i}sica de Canarias, C/ V\'{\i}a L\'{a}ctea, s/n, E-38205 La Laguna, Tenerife, Spain}
\email{msiudek\@iac.es}

\author[0000-0003-1704-0781]{G.~Tarl\'{e}}
\affiliation{University of Michigan, 500 S. State Street, Ann Arbor, MI 48109, USA}
\email{gtarle\@umich.edu}

\author{B.~A.~Weaver}
\affiliation{NSF NOIRLab, 950 N. Cherry Ave., Tucson, AZ 85719, USA}
\email{benjamin.weaver\@noirlab.edu}

\collaboration{all}{(DESI Collaboration)}

\begin{abstract}

We employ a novel data-driven dynamical modeling method, the empirical distribution function (\empdf), to constrain the Milky Way mass profile using a large sample of halo stars from the Dark Energy Spectroscopic Instrument Milky Way Survey (DESI MWS). The method does not assume any functional form of the tracer distribution. Deviations from the steady-state assumption and the effects of observational errors are further tested and calibrated using a number of realistic mock observations (AuriDESI). 
Applying the calibrated method to two classes of tracers in DESI, we derive a best-constrained virial mass and concentration of $\log M_\text{200c}/M_{\odot}=12.05_{-0.12}^{+0.21} \left(_{-0.18}^{+0.26}\right)$ and $\log c_\text{200c}=1.20_{-0.16}^{+0.11}\left(_{-0.24}^{+0.20}\right)$ for BHBs. For RR Lyrae stars, we derive $\log M_\text{200c}/M_{\odot}=12.15_{-0.14}^{+0.22}\left(_{-0.19}^{+0.26}\right)$ and $\log c_\text{200c}=1.16_{-0.12}^{+0.09}\left(_{-0.18}^{+0.16}\right)$. The uncertainties quoted outside and within brackets are statistical and total errors including systematic errors calibrated from AuriDESI, respectively. The two tracer populations provide overall consistent measurements of the MW halo mass profile. Combining the halo star constraints with the independent constraint from satellite dynamics \citet[][\simDF]{Zhaozhou2020}, we derive a robust and precise final measurement of the MW halo parameters of $\log M_\text{200c}/M_{\odot}=12.12_{-0.05}^{+0.05}\left(_{-0.04}^{+0.06}\right)$ and $\log c_\text{200c}=1.16_{-0.03}^{+0.05}\left(_{-0.07}^{+0.05}\right)$, with correlation coefficients of $\rho=-0.90(-0.54)$.  
\end{abstract}


\section{Introduction}

Accurate constraints on the total mass and density distribution of the Milky Way (MW) dark matter halo are fundamental to understanding the formation and evolution of our Galaxy under the cosmological framework. For example, the ``too big to fail'' problem was raised more than ten years ago by \cite{Boylan-Kolchin2011}. They pointed out that the most massive satellite galaxies in our MW are less massive than the most massive subhalos predicted by $\Lambda$CDM simulations of MW-mass systems. This problem, however, can be explained if the virial mass of our MW is smaller than that assumed by \cite{Boylan-Kolchin2011}. For example, \cite{Wang2012} and \cite{Cautun2014} reported that the MW satellites are 
consistent with $\Lambda$CDM predictions at a confidence level of 10\% if the dark halo of the MW has a virial mass smaller than $1.3\times10^{12} M_{\odot}$. This example demonstrates the importance of precisely constraining the MW host halo mass. 

The measurement of the dark matter halo mass of the MW is usually based on the modeling of various types of dynamical tracers, including halo stars, satellite galaxies, cold gas, and even the companion Andromeda galaxy. There are also various types of dynamical methodologies that can be used to model these tracer objects. The readers can refer to \cite{Wang2020} for a review on different methodologies applied to various types of tracer objects, and also to some more recent studies \citep[e.g.][]{WangHF2023,Zhou2023,Jiao2023,Ou2024, Deason2021, Bird2022, Shen2022, Ibata2024, Posti&Helmi2019, Eadie2018,  Wang2022, Callingham2019, Cautun2020, Zhaozhou2020, Kravtsov2024,Roche2024, Li&Han2021}.

Among the different types of methodologies, a common approach for dynamical modeling is based on the phase-space distribution function (DF), in which the tracers are considered to be in dynamical equilibrium, and the model gravitational potential parameters are determined by fitting the model to the observed data \citep[e.g.][]{Zhaozhou2020,Deason2021,Wang2015,Wang2022,Shen2022}. The DF provides a complete description of a dynamical system via its full 6D phase-space representation\citep{BT08}. 

Despite numerous efforts over the past decades devoted to inferring the mass of our MW, the current constraints on the profile of the outer halo show considerable discrepancy \citep{Wang2015,Wang2020}. 
The discrepancy mainly arises from two aspects: 1) limited halo tracer populations with incomplete and inaccurate kinematic data (e.g. proper motions) at large distances; and 2) systematic uncertainties behind the modeling method that remain uncorrected or uncalibrated \citep{Han_opdf1_2016,Han_opdf2_2016,Eadie2018,Deason2021}.
To address these issues and establish reliable constraints on the MW mass profile, improved kinematic observations of halo tracers and reliable modelling methods are essential.

Compared with other types of dynamical tracers (satellite galaxies, globular clusters and stellar streams), halo stars are more numerous, though halo stars stripped from the same progenitor satellite galaxy may carry less independent information than the total sample size due to phase-space correlations \citep[e.g.][]{Wang2017,Wang2018,Han2020}. 

The Milky Way Survey program of the Dark Energy Spectroscopic Instrument \citep[DESI MWS;][]{MWS_overview2023} is observing a substantial sample of stars in the MW outer stellar halo, which can serve as valuable dynamical tracers with line-of-sight velocity measurements. Among these, blue horizontal branch (BHB) stars and RR Lyrae (RRL) variables \footnote{In addition to BHB and RRL, DESI also has a large sample of giant stars observed in the outer stellar halo \citep[e.g.][]{Li2025}, but they are not used in this paper.} are particularly useful due to their nature as distance estimators and their brightness, allowing them to be observed at larger distances from the halo center. The DESI MWS DR2 dataset contains 12,989 BHB and 12,868 RRL stars distributed throughout the halo. This constitutes the largest spectroscopic sample of BHB and RRL tracers assembled to date.
The total number of BHB stars observed by DESI DR1 is almost four times that of previous surveys beyond 50~kpc, with a non-negligible fraction detected beyond 100 kpc \citep{Amanda2024_BHBcat,Medina2025a}.

The large sample of DESI halo stars, especially those probing the more distant outer stellar halo, will provide invaluable new insights into the MW outer halo. By combining DESI line-of-sight velocities and spectrophotometric distances with {\it Gaia} proper motions, these halo stars provide a valuable resource with six-dimensional phase-space information. They can potentially help to constrain the dark matter distribution in the more distant outer stellar halo to better precision, in addition to the use of satellite galaxies as tracers \citep[e.g.][]{Li2019,Zhaozhou2020}. Nevertheless, as we will show in this study, the {\it Gaia} DR3 proper motion measurements are subject to large uncertainties for halo stars beyond 50~kpc, and thus additional corrections or calibrations are required in our measurement.

With the sizable sample of halo stars as tracers, there is a demand for reliable methodologies that can fully exploit the tracer information to estimate the mass of the Galaxy, as well as a good understanding and control of the systematic errors involved. While many DF methods rely on assumed functional forms \citep{McMillan2013,Hattori2021}, a fundamental limitation of such assumed analytical DFs is that the results might be biased if the hypothetical form does not accurately represent the tracer population \citep{Wang2015,Han_opdf2_2016}. 

In response to this limitation of conventional DF methods, a data-driven and minimal assumption method for reconstructing the DF from the data itself under the steady-state assumption was first developed by \citet{Han_opdf1_2016} named the Orbital Probability Distribution function (oPDF), with validations and tests on numerical simulations to understand its underlying systematics \citep{Han_opdf2_2016, Wang2017}. The original oPDF method makes use of the full phase space distribution of tracers to construct a model DF. However, to work around discreteness effects in the model DF, only the predicted radial distribution is compared against observations to infer the best potential model through a likelihood analysis.  
In a recent work, \citet{LiZhaozhou2024} further introduced a smoothing technique to overcome the discreteness in the model DF, which enables the full phase space data to be used when constructing the likelihood. This improved empirical Distribution Function (\empdf hereafter) method is notable for its flexibility, efficiency, and adaptability in managing selection effects, and is about 1.5 times more precise than the traditional Jeans Equation method \citep{LiZhaozhou2024}. It represents a data-driven dynamical modeling method that avoids parametric assumptions on the DF. A related example is the minimum-entropy action-space approach of \citet{Beraldo2025}, which shares similarities with \empdf but stems from a distinct physical motivation. Inheriting from the oPDF method, \empdf does not rely on any pre-assumed functional form of the DF, and can maximize the advantage of using the large sample of DESI halo stars observed so far. 

In this work, we aim to take full advantage of the DESI MWS dataset by applying this data-driven DF method, \empdf, to the BHB and RRL stars from the DESI DR2 observations. We do not only constrain the dark matter distribution around the MW with real DESI halo stars, but also validate the method and calibrate results using mock catalogues presented in \citet{Riley:2026auridesi}. This work is one component of the broader DESI DR2 Milky Way mass Key Project, with complementary mass profile measurements presented in companion papers using other dynamical methods \citep{Medina:2026gme,Valluri:2026nimble}, and an overview and joint interpretation presented in \citet{DESI:2026mwmass}. 

The organization of this paper is as follows. In Section~\ref{sec:data}, we describe the data we used, including both the observational data and mock data. Section~\ref{sec:method} explains how \empdf works. Section~\ref{sec:results} presents the validation and calibration of the method using mock samples, and then apply it to the DESI observations.

\section{ Data}
\label{sec:data}

\subsection{DESI and DESI Milky Way Survey}

DESI is one of the foremost multi-object spectrographs for wide-field surveys \citep{desiScience,desiInstrument,DESI.InstrumentOverview.2022,FocalPlane.Silber.2023,SurveyOps.Schlafly.2023,FiberSystem.Poppett.2024,Spectro.Pipeline.Guy.2023,Corrector.Miller.2024}. 
Millions of Galactic targets have been observed by DESI \citep{Snowmass2013.Levi}, including the Bright Galaxy Survey targets \citep[BGS;][]{BGS.Hahn.2023},
luminous red galaxies \citep[LRG;][]{LRG.Zhou.2023}, emission line galaxies \citep[ELG;][]{ELG.Raichoor.2023}, and quasars \citep{QSO.Chaussidon.2023}. These observations have already produced a wide range of important scientific results, such as measurements of BAO signals from galaxies and the Lyman alpha forest \citep{DESI2024.III.KP4,DESI2024.IV.KP6}, full-shape galaxy clustering \citep{DESI2024.V.KP5}, two-point statistics \citep{DESI2024.II.KP3}, and cosmological constraints \citep{DESI2024.VI.KP7A,DESI2024.VII.KP7B,DESI_DR2_Cosmo}.
 
The DESI MW Survey (MWS) is mainly performed at bright time \citep{MWS_overview2023}, when the sky conditions are bright due to the moon phase, twilight, and worse seeing conditions, making the observation of high redshift galaxies inefficient. The main MW survey targets $\sim 12$ million stars selected from the DESI Legacy imaging Survey \citep{LegacySurveys.Dey.2019} down to a limiting magnitude of $r=19$, with some special types of stars observed down to fainter magnitudes \citep{MWS_overview2023}. It covers most of the northern Galactic cap region and a significant fraction of the southern cap region. 
Beyond observations in bright-time, DESI MWS also has some targets observed in dark-time, including some secondary targets to address specific science goals.

Following the Early Data Release \citet[EDR;][]{DESI2023b.KP1.EDR} of DESI in June 2023, \cite{MWS.VAC.Koposov.2024} published a DESI MWS stellar value-added catalog for 400,000 stars, including commissioning and science validation data\citep{DESI2023a.KP1.SV}. The first data release (DR1; \citealt{DESI_DR1})
is now publicly available, with the corresponding MWS stellar catalogue and
value-added products described by
\citet{DESI.DR1.Stellar.Koposov.2026}.

In this paper, we use the DESI MWS data from its first three years of observation, corresponding to an internal dataset planned for public
release as part of DESI DR2. In particular, our halo star samples are primarily composed of the BHB and RRL stars from the DR2 data.

\subsection{DESI MWS BHB sample}
\label{sec:BHB sample}
The BHB samples used in this study are constructed following the same procedure as \citet{Amanda2024_BHBcat}, but based on the DESI DR2 data. DESI BHB targets are first selected based on photometry and astrometry according to the criteria introduced in \citet[their Section 4.4.4]{MWS_overview2023} and \citet[their Section 2.1]{Amanda2024_BHBcat}. For completeness, the selection criteria are listed below:
\begin{align}
&G>10 \nonumber \\
&\varpi \leq 0. 1+ 3\sigma _\varpi  \mathrm{mas} \nonumber \\
&-0.35\leq(g-r)\leq-0.02 \nonumber \\
&-0.05\leq X_{\mathrm{BHB}}\leq 0.05 \nonumber \\
&r-2.3(g-r)-W1_{\mathrm{faint}}< -1.5 .
\label{eqn:bhbselection}
\end{align}
Here, $G$ is the {\it Gaia} $G$-band magnitude, $\varpi$ is the parallax and $\sigma_{\varpi}$ its uncertainty, while $g$ and $r$ are optical magnitudes from the DESI Legacy Imaging Surveys.

\begin{equation}
\begin{split}
X_{\mathrm{BHB}} = &(g-z)-[1.0716(g-r)^{5}-1.42272(g-r)^{4}\\& + 0.69476(g-r)^{3}-0.12911(g-r)^{2}\\& + 0.66993(g-r)-0.11368],
\end{split}
\end{equation}
represents a polynomial fit to the BHB sequence on the color–magnitude diagram, and
\begin{equation}
W1_{\mathrm{faint}}=22.5-2.5\log_{10}\left(W1-3\sigma_{W1}\right),
\label{eqn:bhbselectionend}
\end{equation}
where $W1$ is the WISE 3.4 $\mu$m flux and $\sigma_{W1}$ is its error. 

As a class of high priority samples in MWS, BHBs are also observed as secondary targets in DESI, which are observed in the dark-time program with magnitude limit $19<r<21$. Although the secondary BHB sample extends to larger Galactocentric distances (up to $\sim120$ kpc), we restrict our analysis to the main survey BHBs observed in bright time, which reach $\sim80$ kpc. The secondary BHBs with $r>20$ constitute only a small fraction of the total sample (2.8\%), but their inclusion would introduce a more complicated selection function. In addition, the currently available mock halo star catalogs (see Section~\ref{sec:mock}) do not provide compatible data products for secondary BHBs in DESI.

After stellar spectra are obtained for the BHB targets, additional cuts are made to remove possible contamination in the final clean BHB sample, as
described in Section~2.4 of \citet{Amanda2024_BHBcat}. Targets spectroscopically classified as quasars are first removed to ensure a purely stellar sample. Then targets with bad quality of radial velocities are also removed according to the quality flags defined by the the DESI MWS radial-velocity
spectroscopy (RVS) pipeline
\citep{RVSpecFit2019,MWS.VAC.Koposov.2024}. Further, contamination from blue stragglers (BS) is also removed. BSs are removed in the $T_{\rm eff}$–$\log g$ (Kiel) diagram, and only stars with 

\begin{equation}
\begin{split}
&7,000<T_{\mathrm{eff}}<14,500 \ \text{K and} \\
&1.50<\log(g)- 0.00014 \ T_{\mathrm{eff}}<2.65
\end{split}
\end{equation}
are retained. Finally, only stars in the color range $-0.3< g-r < 0$ are kept for reliable distances.

\begin{figure*}
\centering
\includegraphics[width=0.7\textwidth]{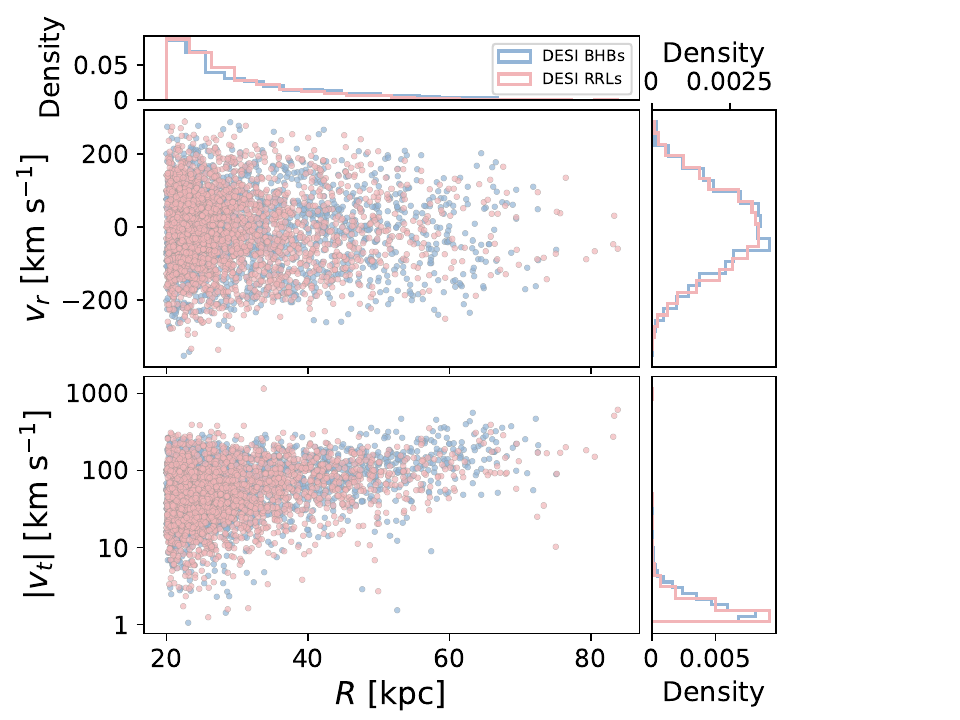}
\caption{ Distribution of Galactocentric distances ($R$), radial velocities ($v_r$), and tangential velocities ($v_t$) for the selected DESI MWS halo stars. Light blue and pink points/histograms represent BHBs and RRLs, respectively. } 
\label{fig:obs distribution}
\end{figure*}

For any mass measurement method based on equilibrium assumptions, the presence of unrelaxed substructures is an important concern. Member stars of the Sagittarius (Sgr) stream have been identified in \citet{Amanda2024_BHBcat} (see their Section~2.6 for detailed selection). In our analysis, we choose to remove the entire Sgr footprint to avoid additional radial selection on the distance. Explicitly, we define boundaries in R.A. and Dec along the Sgr stream and exclude all stars within these boundaries from our analysis. There are 6,687 main bright BHB samples from the catalog after deduplication. After applying the $\texttt{RVS\_WARN}=0$ requirement and removing stars within the Sgr footprint, 4,368 stars remain.\footnote{Our footprint-based removal is more conservative than the distance-dependent Sgr member identification: before the final magnitude and Galactocentric-radius cuts, it removes an additional 830 BHB stars and 822 RRL stars.} Applying the apparent magnitude limit of $16< r_0 < 20$, where $r_0$ is the Galactic extinction corrected $r$-band magnitude, together with the distance cut of $R>20\,\mathrm{kpc}$ yields a final sample of $N=1863$ BHB stars that extend up to $\sim80\,\mathrm{kpc}$ into the stellar halo. The distance and velocity distributions of the final sample are shown in Figure~\ref{fig:obs distribution}.

The line-of-sight velocities of BHBs are given by the DESI MWS RVS pipeline. Proper motions measurements are obtained from {\it Gaia} DR3. Distances were derived from the distance modulus using the absolute magnitudes calculated via the color-dependent relation from \citep{BK15} and \citep{Amanda2024_BHBcat}:

\begin{equation}
M_g = 
\begin{cases} 
\begin{split}
& 0.566 - 0.392(g-r) + 2.729(g-r)^2 \\
& \quad + 29.1128(g-r)^3 + 113.569(g-r)^4,
\end{split} & \\ 
\text{for } -0.30 < (g-r) < -0.05, \\
0.564 - 0.500(g-r), & \\
\text{for } -0.05 < (g-r) < 0.00.

\end{cases}
\end{equation}
In this work, Heliocentric quantities are converted into Galactocentric ones using the \texttt{Galactocentric} frame in \textsc{Astropy}\citep{Astropy:2013,Astropy:2018,Astropy:2022}.

\subsection{DESI MWS RR Lyrae sample}

The DESI MWS RRL targets are selected from {\it Gaia} Data Release 2 \citep[DR2;][]{Clementini2019} and the Pan-STARRS1 \citep[PS-1;][]{Sesar2017} catalogs. Details about the RRL sample can be found in \cite{Medina2025a,Medina2025b}. The DR2 RRL sample consists of stars observed by DESI as part of the main and the Survey Validation (SV) programs, crossmatched with the {\it Gaia} DR3 RRL catalog \citep{GaiaDR3_2023}, with G-band absolute magnitudes computed from the luminosity-metallicity relation calibrated by \citet{Garofalo2022}:
\begin{equation}
M_G = (0.28_{-0.36}^{+0.36})[\rm{Fe/H}] + (0.97_{-0.52}^{+0.49}).
\end{equation}
The radial pulsating velocities of DESI RRLs are corrected following the method of \cite{Medina2025a}. 

Similar to BHB stars, we also remove the Sgr footprint from our sample of RRL, leaving 4,838 stars. Applying the same apparent magnitude limit of $m_{\rm lim}=20$ in addition to the distance cut of $R>20$ kpc, we obtain a final sample of 2,282 RRL stars with Galactocentric distances extending to about 80 kpc.

\subsection{AuriDESI mock sample}
\label{sec:mock}

For our calibration and validation, we use the DR2 AuriDESI BHB and RRL mock catalogues described in \citet{Riley:2026auridesi}. The mock halo star catalog is constructed from the cosmological Auriga simulations \citep{Grand2017} at level-3 resolution. It follows the particle sampling approach of \cite{Lowing2015} to decompose each star particle in the parent hydrodynamical simulation into individual stars. The stellar parameters of each individual star are obtained by matching the star particle to the closest PARSEC isochrone \citep{2012MNRAS.427..127B} according to its stellar age and metallicity.

The mock DESI halo star catalogs used in this study are based on the AuriDESI suite \citep{AuriDESI2024}, a new set of mocks specifically constructed for DESI observations; see \citet{Riley:2026auridesi} for a detailed description of the DR2 BHB and RRL mock catalogs. These mocks provide mock realizations of the DESI MWS dataset based on the previous AuriGaia ICC-MOCKS \citep{2018MNRAS.481.1726G}, which themselves are derived from the Auriga simulation suite. 

There are six MW-like systems in total, each of which includes four sets of catalogs with different viewing directions, corresponding to four different positions of the observer, located in the disc plane of the simulation and along four different directions ($30^{\circ}$, $120^{\circ}$, $210^{\circ}$, $300^{\circ}$), with respect to the bar major axis. Apparent magnitudes and sky coordinates are computed for each observer position, and the DESI survey flux limits and footprint masks are then applied. The AuriDESI catalogs provide flags identifying stars associated with bound satellite galaxies and stellar streams \citep{Riley2025}. We exclude both components from the mock samples in order to construct tracer catalogs that more closely approximate dynamical equilibrium. 

To select BHBs and RRLs from the mock stellar catalog above, the following criteria are applied to first define the horizontal branch (HB), including cuts in Kiel space with $ 5500 < T_{\rm{eff}}/\rm{K} < 9750 $ and  $0.7765 + T_\text{eff}/3863 < \log g < 1.4059 + T_\text{eff}/4250$, an absolute magnitude limit of $0.4 < M_g < 0.85 $, and an age selection of Age > 8 Gyr. BHBs are then defined as stars along the HB with $T_{\mathrm{eff}}>7000~\mathrm{K}$, while RRLs are those with $6000<T_{\mathrm{eff}}<7000~\mathrm{K}$. These selections are made based on the actual stellar parameters in AuriDESI, obtained from the PARSEC isochrones. 

\begin{figure*}
\centering
\includegraphics[width=1.99\columnwidth]{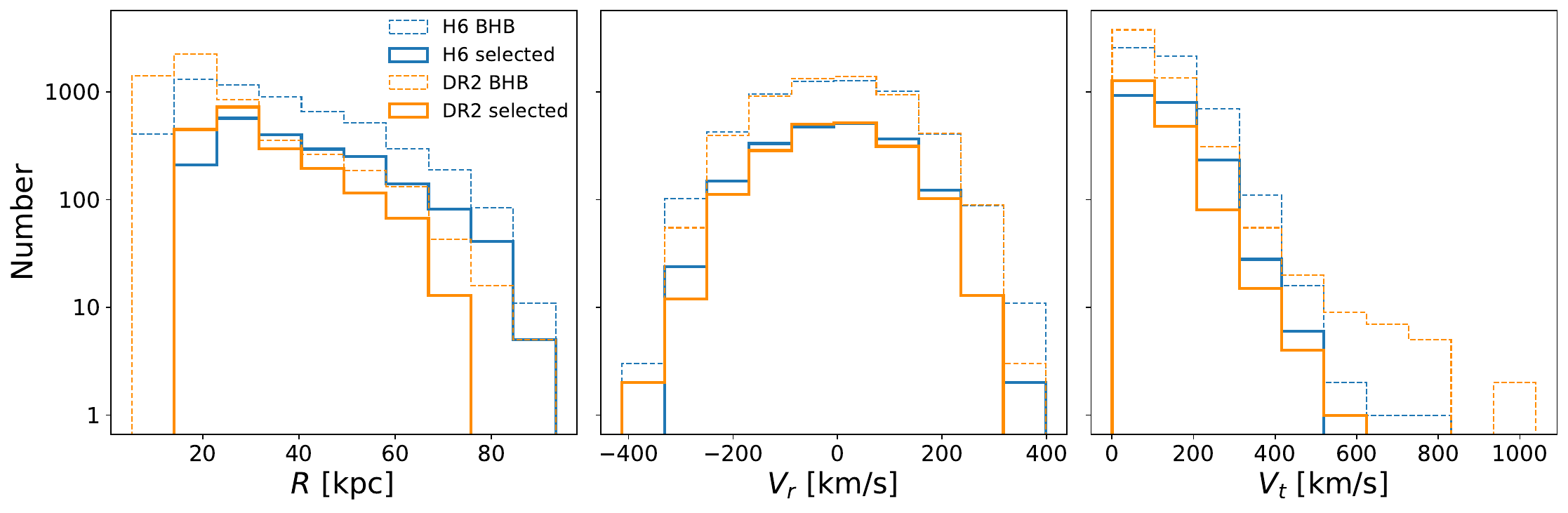}
\caption{ Distribution of Galactocentric distance ($R$), radial velocity ($v_r$), and tangential velocity ($v_t$) for the mock and observed BHBs in DR2. Blue histograms show mock data distribution based on AuriDESI halo H6 with the mock observer at a viewing direction $30^{\circ}$ from the galactic bar axis. Only H6 is shown here for clarity, as other halos exhibit similar distributions. The solid line histograms represent the final used sample, while dashed histograms correspond to the sample before selection (see Section~\ref{sec:BHB sample}). } 
\label{fig:obs vs mock distribution}
\end{figure*}

With the mock BHB and RRL samples constructed above, the survey selection functions are further incorporated through the AuriDESI observational probability. The DESI spectroscopic completeness component follows the selection-function construction described in Section~2.4 of \citet{DESI:2026mwmass}. This component describes the completeness fraction of spectroscopically observed DESI BHBs and RRLs relative to the original photometric targets, which is a function of the number of passes/exposures and Galactic coordinates, with weak dependence on apparent magnitudes or stellar color (we will discuss this in Section~\ref{sec:selectfunc} below).  The observational probability is then applied to the mock BHB and RRL samples by randomly discarding stars at each sky pointing, such that the mock completeness matches that of the real DESI observations.
For consistency with the observational samples, we apply the same magnitude cuts of $16<r_0<20$ for BHBs and $16<G<20$ for RRLs to the mock samples. We also impose an inner Galactocentric radius cut of $20\,\mathrm{kpc}$. To ensure that the final numbers of mock BHB and RRL tracers match those in the real DR2 samples, further downsampling is applied to the selected mock catalogs. After this step, each mock realization contains approximately 2,000 BHB stars and 2,300 RRL stars, comparable to the corresponding observational samples. 
The resulting distributions of distances and velocities in the mock BHB sample broadly cover similar radial and kinematic ranges, although the AuriDESI tracers are less centrally concentrated than the DESI data, as shown in Figure~\ref{fig:obs vs mock distribution}.

To account for observational effects, we have two types of mock samples: 1) error-free samples where the coordinates and velocities are the actual values from the original mock; 2) error-added samples, which consist of 20 DESI-like observational realizations for each halo and viewing angle for each tracer. Comparing the results from the error-free and error-added samples allows us to assess the impact of observational errors on the model inference. 

For the error-added realizations, the observed distances are perturbed by adding Gaussian noise with a standard deviation equal to $5\%$ of the distance, consistent with typical distance uncertainties for DESI BHBs and RRLs. The line-of-sight velocity uncertainty of  BHBs is assigned from the DESI DR2 BHB error model in the $(g-r,r)$ plane. For RRLs, the uncertainty on the observed line-of-sight velocity is drawn directly from the empirical uncertainty distribution of the DESI DR2 RRL catalogue. 
The proper motion errors, which dominate the uncertainty on the 3-D velocity, are based on the {\it Gaia} DR3 error model and increase with apparent magnitude. For each halo, we define the reference NFW parameters used throughout the mock tests by fitting a spherical NFW profile to the true spherically averaged total enclosed-mass profile over 20 to 150 kpc.

\section{Method}
\label{sec:method}

\subsection{\empdf}

In this work, we adopt the Empirical Distribution Function (\empdf) method developed by \citet{LiZhaozhou2024}. This method allows us to infer the underlying potential from the observed distribution of tracers, without any assumption about the functional form of the DF. It is a minimal assumption method, with the only essential assumption being that the system is in a steady state. Here, we also make an assumption of a spherically symmetric underlying potential. 

The main concept behind \empdf is that, in a steady state system, the underlying DF of tracers remains time-invariant. Therefore, the degree of similarity between the instantaneous observed kinematics of tracers and the time-averaged statistics empirically reconstructed under a trial potential reflects how well the model fits the data.

\empdf is an inheritance and extension of the orbital Probability Density Function (\opdf) method introduced by \citet{Han_opdf1_2016}, which states that the probability of observing a tracer at a given position is proportional to the time it spends at that position in a steady-state system, $\mathrm{d}P(\lambda|\mathrm{orbit})\propto\mathrm{~d}t$, where $\lambda$ represents the position in the orbit.
In a spherical potential, the distribution along a given orbit can be expressed via a radial coordinate as $p(r | E,L)dr\propto dt=dr/|v_r|$. Here, $E$ and $L$ denote the specific orbital energy and the angular momentum, respectively, while $v_r$ is the radial velocity. Normalizing over a radial orbital period $T_r$ yields
\begin{equation}
p(r | E,L)=\frac{1}{|v_r|T_r}.
\end{equation}
This conditional distribution function is equivalent to time-averaging the radial distribution of each particle along its orbit.

Assuming spherical symmetry, we can decompose the DF $f(\mathbf{r,v})$ into the distribution of orbits, $p(E,L)$, and the radial distribution, $p(r|E,L)$, along the orbit,
\begin{equation}
f(\mathbf{r,v})=\frac{|v_r|}{8\pi^2L}p(r|E,L)p(E,L).
\end{equation} 

A smooth orbital distribution $p(E,L)$ in \empdf is approximated by kernel density estimation \citep[KDE;][]{Silverman1986_KDE},
\begin{equation}
p(E,L)=\frac{\sum_{i=1}^N\omega_i\mathcal{K}(E-E_i,L-L_i)}{\sum_{i=1}^N\omega_i},
\label{eq:9}
\end{equation}
where $\omega_i$ is the weight of each tracer, and $\mathcal{K}$ is the Epanechnikov kernel,  $\mathcal{K}(\boldsymbol{u})=2(1-\|\boldsymbol{u}\|^2)/\pi$ for $\|\boldsymbol{u}\|<1$ and zero otherwise, where $\boldsymbol{u}$ denotes the bandwidth-scaled separation in the orbital-parameter space. The bandwidth of the kernel is determined using Scott's rule \citep{SCOTT1979}.
Combining the conditional radial distribution along each orbit with the continuous orbital distribution, a continuous, data-driven steady-state DF is constructed as 
\begin{equation}
f_\Phi(\mathbf{r,v})=\frac{p(E,L)}{8\pi^2LT_r(E,L)},
\end{equation} 
facilitating direct comparison with observations. 
Bayesian inference can be applied to derive the correct potential since the likelihood between the empirical DF and the observations is computable.

The probability of observing a tracer with $\boldsymbol{w}_i\equiv(\boldsymbol{r}_i,\boldsymbol{v}_i)$ under a trial potential, $\Phi(r)$, specified by parameters $\boldsymbol{\Theta}_\Phi$ is
\begin{equation}
    p(\boldsymbol w_i|\boldsymbol{\Theta}_\Phi)=f_\Phi(\boldsymbol{w_i}).
\end{equation}
The posterior of the model potential parameters for this tracer sample is
\begin{equation}
    p(\boldsymbol{\Theta}_\Phi|\{\boldsymbol{w}_i\})\propto p(\boldsymbol{\Theta}_\Phi)\prod_{i=1}^Np(\boldsymbol{w}_i|\Phi),
\end{equation}
where $p(\boldsymbol{\Theta}_\Phi)$ is the prior on the potential parameters. By maximizing the posterior, the best-fit potential can be obtained.

In this study, we adopt the assumption that the total mass distribution of the MW follows a spherical NFW profile \citep{NFW1996}, characterized by two parameters: the virial mass $M_{200\mathrm{c}}$, and the concentration $c_{200\mathrm{c}}$. $M_{200\mathrm{c}}$ is defined as the total mass enclosed within a radius $r_{200\mathrm{c}}$, where the average density is 200 times the critical density of the universe, and $c_{200\mathrm{c}}$ is defined as the ratio of the virial radius $r_{200\mathrm{c}}$ to the scale radius $r_{\mathrm{s}}$ in the NFW density profile. For simplicity, we will hereafter refer to these parameters as $M$ and $c$, respectively. The likelihood is estimated by 2-dimensional grid scanning over this parameter space. In the default setup used for the mock samples and for the DR2 BHB sample, we adopt $\log M_{200\mathrm{c}}/M_\odot\in[11.5,13.0]$ and $\log c_{200\mathrm{c}}\in[0,2]$. For the DR2 RRL sample, we extend the mass range to $[11.5,13.5]$ to ensure that the posterior is fully enclosed by the scanned range.

Due to the inability of a single NFW profile to adequately describe the complex stellar components in the central region, including the stellar disk and other aspherical structures, our analysis is restricted to $r > 20$ kpc. At these radii, the baryonic contribution is reduced, and hydrodynamical simulations suggest that the total mass profiles of MW-sized halos can be reasonably approximated by an NFW profile \citep {Schaller2015,Kelley2019}.

\subsection{Survey selection function correction}
\label{sec:selectfunc}

\begin{figure}

\centering

\includegraphics[width=0.96\columnwidth]{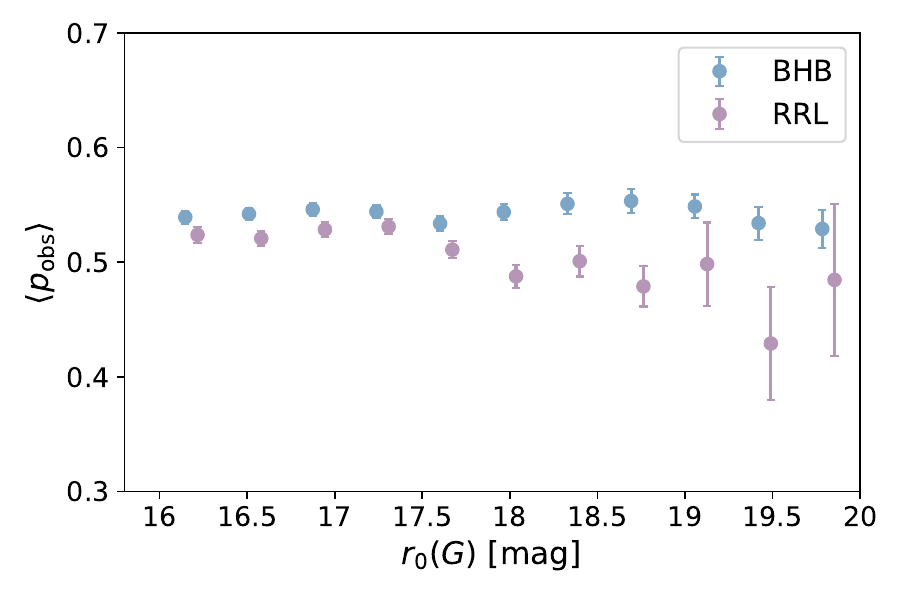}
\caption{Mean detection probability, $\langle p_{\rm obs}\rangle$, as a function of apparent magnitude for the final BHB and RRL samples used in the \empdf analysis. Blue points with error bars show $\langle p_{\rm obs}\rangle$ as a function of Galactic extinction corrected $r$-band apparent magnitude, $r_0$, for BHBs. Violet points show $\langle p_{\rm obs}\rangle$ as a function of {\it Gaia} $G$-band apparent magnitude for RRLs. Points represent the mean value in each magnitude bin, and error bars show the standard error of the mean. For BHBs, $p_{\rm obs}$ mainly reflects the DESI spectroscopic completeness, while for RRLs it also includes the {\it Gaia} RRL target catalogue completeness.}
\label{fig: mean_pobs}
\end{figure}

In our analysis, we do not explicitly correct for the angular selection function or the dependence of the completeness fraction on sky coordinates, because we assume spherical symmetry in \empdf. The observational catalogues provide a detection probability, $p_{\rm obs}$, for each star. To check whether the final tracer samples used in our analysis show a significant magnitude dependence in this observational probability, we calculate the mean $\langle p_{\rm obs}\rangle$ in bins of apparent magnitude. As shown in Figure~\ref{fig: mean_pobs}, $\langle p_{\rm obs}\rangle$ is nearly independent of apparent magnitude over the adopted flux-limited range. For BHBs, $p_{\rm obs}$ mainly reflects the DESI spectroscopic completeness, while for RRLs it also includes the {\it Gaia} RRL target catalogue completeness. Therefore, within the adopted flux-limited range of $16<r_0<20$ for BHBs and $16<G<20$ for RRLs, which is consistently applied to both the observational and mock samples, we do not apply any additional treatment of the selection function in the flux-limit range.

Although the observational probability does not show a significant dependence on magnitude within the adopted range, the flux limit of the DESI MWS program excludes BHB and RRL stars at certain distances. Because BHB and RRL lie on the horizontal branch in the color-magnitude diagram with nearly constant absolute magnitudes, the distance modulus allows this flux limit to define a distance window, which has to be properly considered in the modeling. Explicitly, this radial selection function is corrected through the following two aspects:

1. Each tracer is a biased representative of its orbital parameters $(E_i,L_i)$. To account for the missed tracers, we reweight each observed tracer with these calculated weights when computing $p(E,L)$ in Equation~\ref{eq:9},
\begin{equation}
\omega_i=\frac{\int p(\boldsymbol{w}^{\prime}|E_i,L_i)\mathrm{d}^6\boldsymbol{w}^{\prime}}{\int p(\boldsymbol{w}^{\prime}|E_i,L_i)S(\boldsymbol{w}^{\prime})\mathrm{d}^6\boldsymbol{w}^{\prime}} \geq 1,
\end{equation}
when constructing the empirical orbital distribution. Here $S(\boldsymbol{w}) \in [0,1],$ denotes the selection function, which quantifies the probability that a tracer at phase-space position is included in the observed sample. In many practical cases, $S(\boldsymbol{w})$ acts as a binary window function determined by survey limits. In our case, this is precisely the situation: the flux limit of the star samples defines a sharp observable distance window for each star, leading to $S(\boldsymbol{w})=1$ if the star's distance falls within its window and $S(\boldsymbol{w})=0$ otherwise (see Section~2.3 in \citet{LiZhaozhou2024} for details). 

2. An observed tracer sample follows the underlying DF modulated by the selection function. We therefore need to renormalize the selected DF before evaluating the likelihood:
\begin{equation}
    p(\boldsymbol{w}_i|\Theta_\Phi)=\frac{f_\Phi(\boldsymbol{w}_i)S(\boldsymbol{w}_i)}{\int f_\Phi(\boldsymbol{w}^{\prime})S(\boldsymbol{w}^{\prime})\mathrm{d}^6\boldsymbol{w}^{\prime}}.
\end{equation}

When considering the selection function correction above, as illustrated in \cite{LiZhaozhou2024}, in principle, Heliocentric distance instead of the Galactocentric radius should be used. Yet here we neglect the offset between the Sun and the Galactic center, since this effect is minimal for stars beyond 20~kpc (see Section~\ref{sec:data}).

In our analysis throughout this paper, the same method of correcting for survey selection function is applied to both real observed BHBs/RRLs and to mock BHBs/RRLs.

\subsection{Estimator bias and systematic uncertainties}~\label{sec:bias}

Real data is subject to observational errors. 
Generally, the observational errors will spread and bias the distribution of the true coordinate, $\boldsymbol{w}_{\mathrm{tr}},$ to the observed coordinate, $\boldsymbol{w}$, which may result in a bias in our best constrained potential model or mass profiles. 
Similar to the selection function, observational errors affect \empdf in two aspects: 
the construction of the empirical DF and the evaluation of the likelihood under this DF.
The current implementation of \empdf does not explicitly model observational errors. This is not an issue in \citet{LiZhaozhou2024}, where systematic uncertainties from observational errors are negligible when using bright satellite galaxies and globular clusters as tracers, since their phase-space coordinates have significantly smaller uncertainties than those of individual halo stars.
However, this assumption does not hold when employing halo stars as dynamical tracers, where the larger observational errors of halo stars in our analysis could substantially affect the results. As we will show in Section~\ref{sec: r020 mock} using mock observations, in the presence of observational error, \empdf tends to overestimate the slope of the mass profile.

In addition to observational error, other deviations from the model assumptions in \empdf could also introduce systematic biases and uncertainties. Due to the minimal assumption nature of \empdf, the remaining deviations from our model assumptions include deviations from steady-state, spherical symmetry, and the adopted NFW form of the density profile. A number of previous works using \opdf have shown that the net effect of these deviations, primarily due to the deviation from steady-state, is to produce a stochastic bias of individual estimates, thereby increasing the overall uncertainty of the inference, while the averaged best-fit over an ensemble of systems remains unbiased (ensemble-unbiased)~\citep{Han_opdf2_2016,Wang2017,Wang2018,QYLi2021}. As we will show in section~\ref{sec: r020 mock}, this conclusion also holds for \empdf, which is an extension to the \opdf method.

Accounting for biases and systematic uncertainties, the resultant $x_{\rm fit}$ can be written as: 
\begin{equation}
 x_{\rm fit} = x_{\rm true} + \Delta_x + \epsilon_{\rm stat} + \epsilon_{\rm sys},\label{eq:fit_decompose}
\end{equation}
where $x_{\rm fit}$ is the best-fit parameter estimated by our model. The term $\Delta_x = \langle x_{\rm{fit}} - x_{\rm{true}} \rangle$ represents the net systematic bias derived from our mock analysis, as will be shown in section~\ref{sec: r020 mock}. The uncertainty terms are assumed to be normally distributed with $\epsilon_{\mathrm{stat}} \sim \mathcal{N}(0, \sigma_{\mathrm{stat}}^2),\  \epsilon_{\mathrm{sys}} \sim \mathcal{N}(0, \sigma_{\mathrm{sys}}^2)$, where $\sigma_{\rm sys}$ and $\sigma_{\rm stat}$ represent the systematic and statistical uncertainties, respectively. 
Throughout this paper, unless otherwise specified, we use the term ``statistical uncertainty'' to refer to the formal statistical uncertainty inferred from the \empdf posterior. The underestimation of this statistical uncertainty, which mainly includes the effects of observational error, together with all other systematic uncertainties due to deviations from model assumptions, is incorporated into the $\sigma_{\rm sys}$ term. 

In our analysis, the systematic uncertainty is specified with a two-dimensional covariance matrix for $M$ and $c$, and the total uncertainty is derived by convolving the posterior distribution of \empdf with the systematic covariance. This convolution is not easy to implement if we only provide our results in the form of Equation~\eqref{eq:fit_decompose}. To avoid this complication, in our final results we will directly quote the total uncertainty instead of the systematic one. Our final reported parameters will be in the form of
\begin{equation}
    x = \hat{x} \pm \sigma_{\rm stat} (\pm \sigma_{\rm tot}),
\end{equation} where $\hat{x}=x_{\rm fit}-\Delta_x$ is the bias-corrected best-fit value (with $\Delta_x$ applied, as calibrated in section~\ref{sec: r020 mock}).

\section{ Results}
\label{sec:results}

\subsection{Tests with mock samples}
\label{sec: r020 mock}

In this subsection, we present our best constrained mass profiles and halo parameters based on the mock sample, and compare with the truth. We mainly present results based on the BHB mock, as the RRL mock follows the same qualitative trends and leads to similar conclusions.

\subsubsection{Calibration of biases}
\label{sec:mock-cal}

Figure~\ref{fig:r0<20 BHB err-free VS err-added mass profile} compares the best-fit mass profiles to the true profiles, using mock BHB samples with (blue curves) and without (green curves) observational errors. The top six panels refer to different halos (see the legend) viewed from the same direction. The bottom panels show the results for one of the halos (H27) viewed from three other directions, and the results for other halos are qualitatively similar. 
This figure allows us to evaluate and compare the results both across different halos observed from the same viewing direction and for the same halo viewed from different directions. From these plots, we find that without observational errors, the best constraints (green curves) show relatively smaller deviations from the truth (black curves) in each panel. The green curves are about equally likely to lie above or below the true profiles. 

In the left panel of Fig.~\ref{fig: r020 err-added bestfit total} we show the distribution of best-fit parameters for all the error-free mock BHB samples. The mean of the distribution as marked by the black cross appears slightly biased towards a lower mass and a higher concentration. Possible contributors may include the non-uniform survey footprint and halo substructure, and the nature of the selected old, metal-poor tracers. However, given the small number of systems used in this analysis and the correlations among the data points from the same halo, the uncertainty on the mean is likely underestimated, and we avoid over-interpreting this small bias. Previous works using a much larger number of independent halos have clearly shown that the result from oPDF, the precursor of \empdf, is ensemble unbiased~\citep{Han_opdf2_2016,Wang2017,QYLi2021}. So we conclude that the results here are also consistent with being ensemble unbiased. Moreover, any residual bias due to the \empdf estimator itself will also be accounted for when we calibrate the result using the more realistic error-added mocks.

In contrast to the results from error-free mocks, the inclusion of observational error leads to a significant overall bias, represented as a steeper slope and higher mass at larger radii of the mass profile (blue curves with hatched bands in Figure~\ref{fig:r0<20 BHB err-free VS err-added mass profile}) than the truth (black). The bias mainly arises from the fact that observational errors increase rapidly with distance. In particular, the proper motion errors contribute a significant amount of uncertainty as the distance increases, as shown in Figure 2 in \citet{Bird2021}. 
The increasing observational errors with distance inflate the observed velocity dispersion, hence biasing the best constrained mass higher in the outer halo.

This bias is also clearly visible in the distribution of the best-fit parameters in the right panel of Fig.~\ref{fig: r020 err-added bestfit total}, as an overestimation of the mass and an underestimation of the concentration for the majority of mock halos. The average biases combining all the mock samples are listed in Table~\ref{tab:mock calibration}. 
We adopt these average bias values to calibrate our measurements based on real DESI halo stars. Note this bias accounts for total bias in fitting \empdf to a realistic mock observation, which is mainly driven by observational error but also includes residual biases from any other sources of systematics that are present in the mock.

\subsubsection{Calibration of formal and total uncertainties}
\begin{figure*}
\centering
\includegraphics[width=1.4\columnwidth]{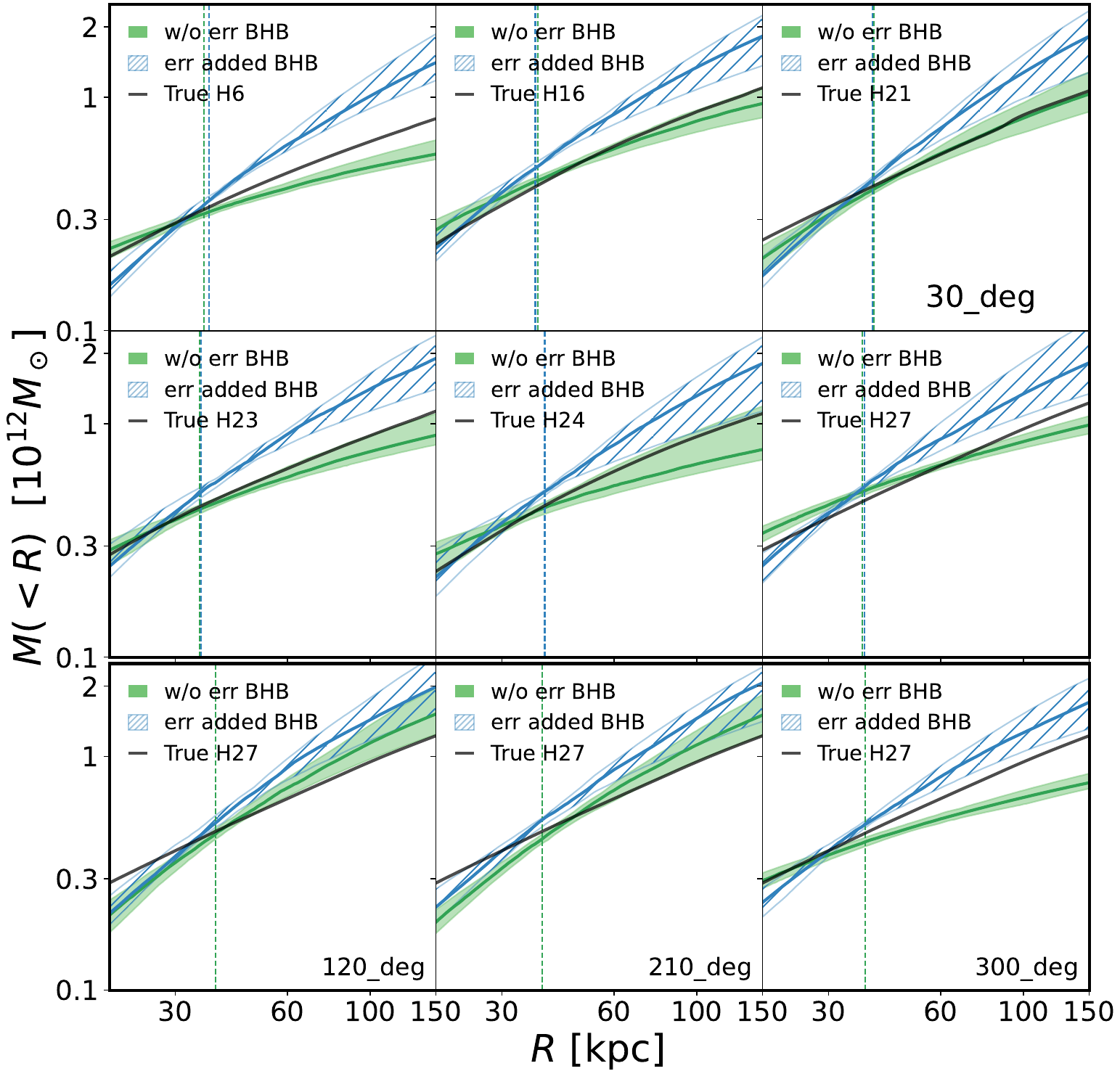}  
\caption{Comparison of the enclosed mass profiles derived from error-free and error-added mock BHB samples. In each panel, the black curve shows the true mass profile of the halo, and the vertical dotted line represents the median radius of tracers. The green curve shows the best-fit mass profile from error-free mock sample and associated shaded band is the 1$\sigma$ confidence region. The blue curve and hatched band correspond to the mean and 1$\sigma$ scatter of the best-fits of 20 error-added realizations for each halo and viewing angle. The top six subpanels show results for different halos from the same observing direction (30$^\circ$). The three bottom panels show results for the same halo (H27) from the remaining directions.} 
\label{fig:r0<20 BHB err-free VS err-added mass profile}
\end{figure*}

\begin{figure*}
\centering
 \includegraphics[width=0.96\columnwidth]{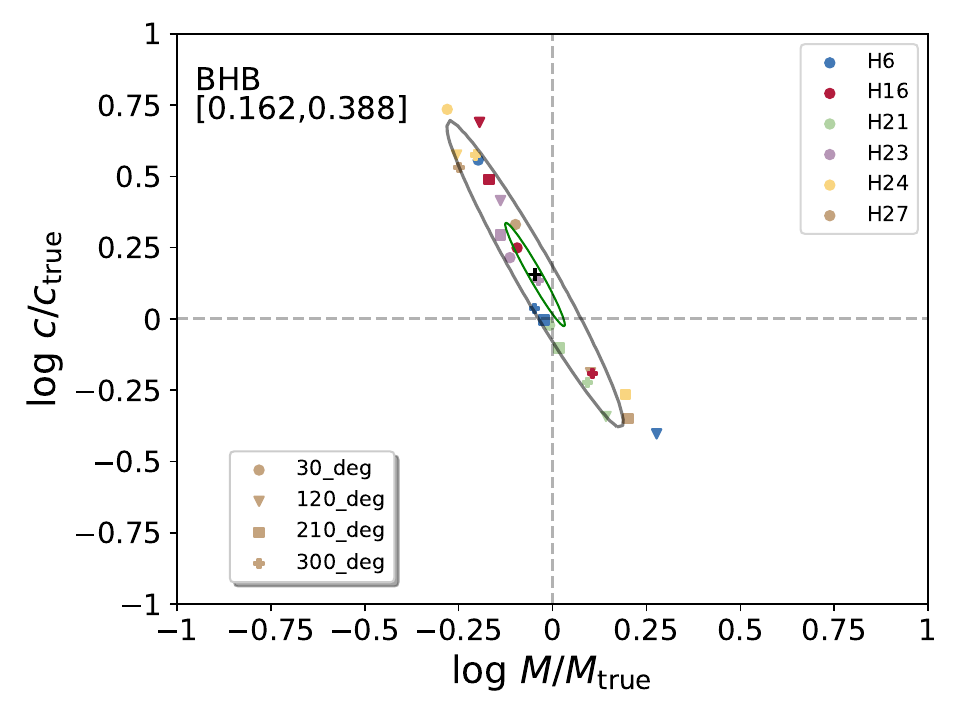}  
\includegraphics[width=0.96\columnwidth]{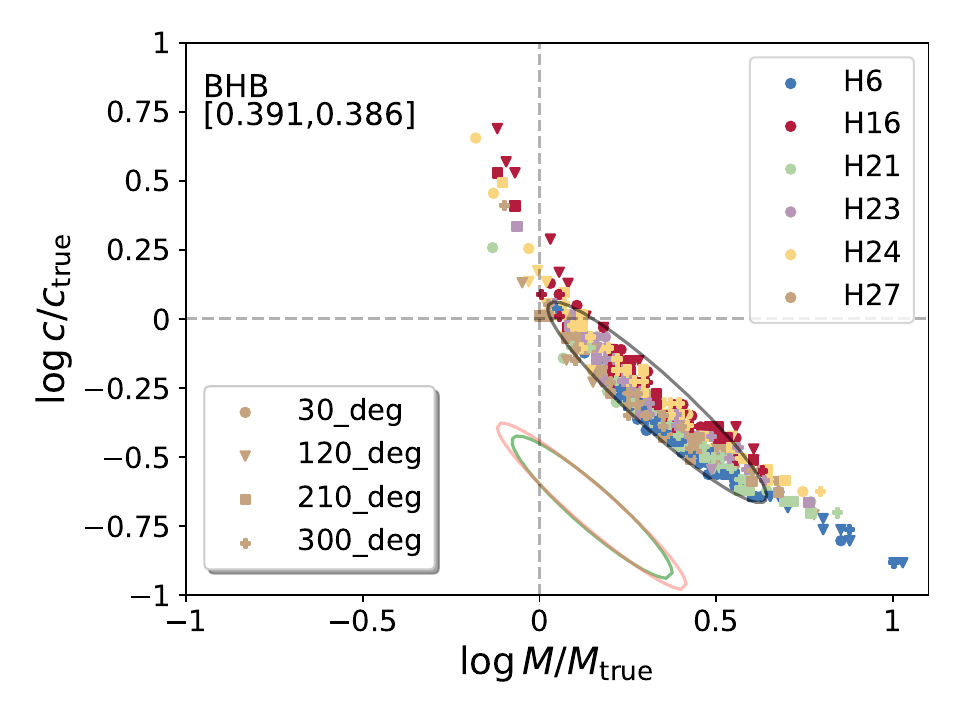}  
\caption{The best-fit halo mass ($M$, $x$-axis) and concentration ($c$, $y$-axis) versus the true parameters for all halos and all viewing directions. Different colors denote different systems, and different symbols mark the four viewing directions (see legend).  {\bf Left:} Application of \empdf to error-free mock data. The black cross and green ellipse represent the mean and uncertainty of all symbols. The best constraints are approximately unbiased after ensemble average.  {\bf Right:} Similar to the left plot, but showing the results of applying \empdf to the mock data with realistic observational errors incorporated. For each halo and viewing direction, there are 20 realizations. The green ellipse represents the average size of $1\sigma$ formal error contour from the \empdf output of each realization. The pink ellipse represents the average size of $1\sigma$ statistical error contour of each 20 realizations. In both plots, the gray ellipses mark the $1\sigma$ total scatter of different systems (all halos, directions, and realizations). The root-mean-squared error relative to true values are illustrated in the bracket on the top left corner.}
\label{fig: r020 err-added bestfit total}
\end{figure*}

Figure~\ref{fig:r0<20 BHB err-free VS err-added mass profile} also shows that the formal error on the mass profile derived from \empdf is slightly lower than the standard deviation among different realizations when observational errors are added to the same halo. 
This is also seen in the right panel of Fig.~\ref{fig: r020 err-added bestfit total}, where the formal posterior contours is comparable to the scatter among realizations, although the latter is slightly larger. The scatter among error-added realizations provides an empirical estimate of the single-halo, single-view statistical scatter. The formal posterior uncertainty primarily reflects the finite information content of the tracer sample under the adopted likelihood.
In addition, the total scatter among the best-fits from all the mocks is larger than the formal or statistical error derived for a single halo. As discussed in Section~\ref{sec:bias}, this enlarged uncertainty reflects additional stochastic biases that are not accounted for by the net bias calibration, $\Delta$, including both deviations from the model assumptions and effects of observational errors not captured by the formal uncertainty. We quote the formal uncertainty as the representative statistical uncertainty throughout this paper, instead of introducing an additional term alone. Since the additional excess scatter is incorporated into the total uncertainty.

Subtracting the average formal covariance matrix from the total covariance measured from all the error-added mocks, we obtain the residual systematic covariance listed in Table~\ref{tab:mock calibration}. In our subsequent analysis using real DESI observations, we convolve the posterior distribution of model parameters produced by \empdf with a Gaussian distribution characterized by this residual systematic covariance matrix to account for the systematic uncertainty. With this construction, the quantity reported as $\sigma_{\rm tot}$ below represents the uncertainty after convolving the formal posterior derived from \empdf with the residual systematic covariance calibrated from the mock ensemble.

Since we have six halos observed in 4 different directions, there are 24 different systems in total. Figure~\ref{fig: r020 err-added bestfit total} provides an overall uncertainty, which actually contains direction-to-direction and halo-to-halo scatters. In Appendix~\ref{app:sys_component}, we discuss the contributions from these two sources separately. The analysis demonstrates that the halo-to-halo scatter is larger than the direction-to-direction scatter, indicating that variations among halos dominate the systematic uncertainties.

We also perform the same analysis using mock catalogs of RRL and obtain similar conclusions, with the calibrated systematics listed in Table~\ref{tab:mock calibration}. For brevity, we do not repeatedly show the results for mock RRLs.

\begin{table}[htb]
\centering
\renewcommand{\arraystretch}{1.3}
\caption{Systematic bias and systematic error covariance matrix of the best-fit parameters, calibrated using error-added mock BHBs and RRLs (see Section~\ref{sec:mock-cal}). The second and third columns show the biases of [$\Delta\log M, \Delta \log c$] and the systematic covariance error matrix formed by $\log M$ and $\log c$. }
\begin{tabular}{lcc}
\hline
Tracer & Bias  & Systematic covariance \\
\hline
BHB &[ 0.333, -0.302 ]& $\bigl[\begin{smallmatrix}
 0.0194 & -0.0223 \\
-0.0223 &  0.0283
\end{smallmatrix}\bigr]$\\
\hline
RRL & $[ 0.579, -0.463]$ & $\bigl[\begin{smallmatrix}
  0.0159 & -0.015 \\
-0.015&  0.0167
\end{smallmatrix}\bigr]$ \\
\hline
\end{tabular}
\label{tab:mock calibration}
\end{table}

\subsubsection{Best constraint at the median tracer radius}

As seen in Figure ~\ref{fig:r0<20 BHB err-free VS err-added mass profile}, the enclosed mass is most tightly constrained near the median radius of the tracers, $r_{\rm med}$, where both the scatter and bias reach their minimum. Such a characteristic radius generally exists in the results from various types of dynamical models \citep{Wolf2010,Han_opdf1_2016,Wang2017,Li2019,QYLi2021,LiZhaozhou2024,Shi2024}. It can be intuitively understood as that dynamical tracers mostly constrain the gravitational acceleration at a representative tracer radius.
In addition to the uncertainty in the mass-concentration plane, it is also informative to examine the uncertainty of enclosed mass within the half-mass or median radius of tracers, which can serve as a concise summary of the highly correlated uncertainties in the mass and concentration parameters. 
 
\begin{figure*}
\centering
\includegraphics[width=\columnwidth]{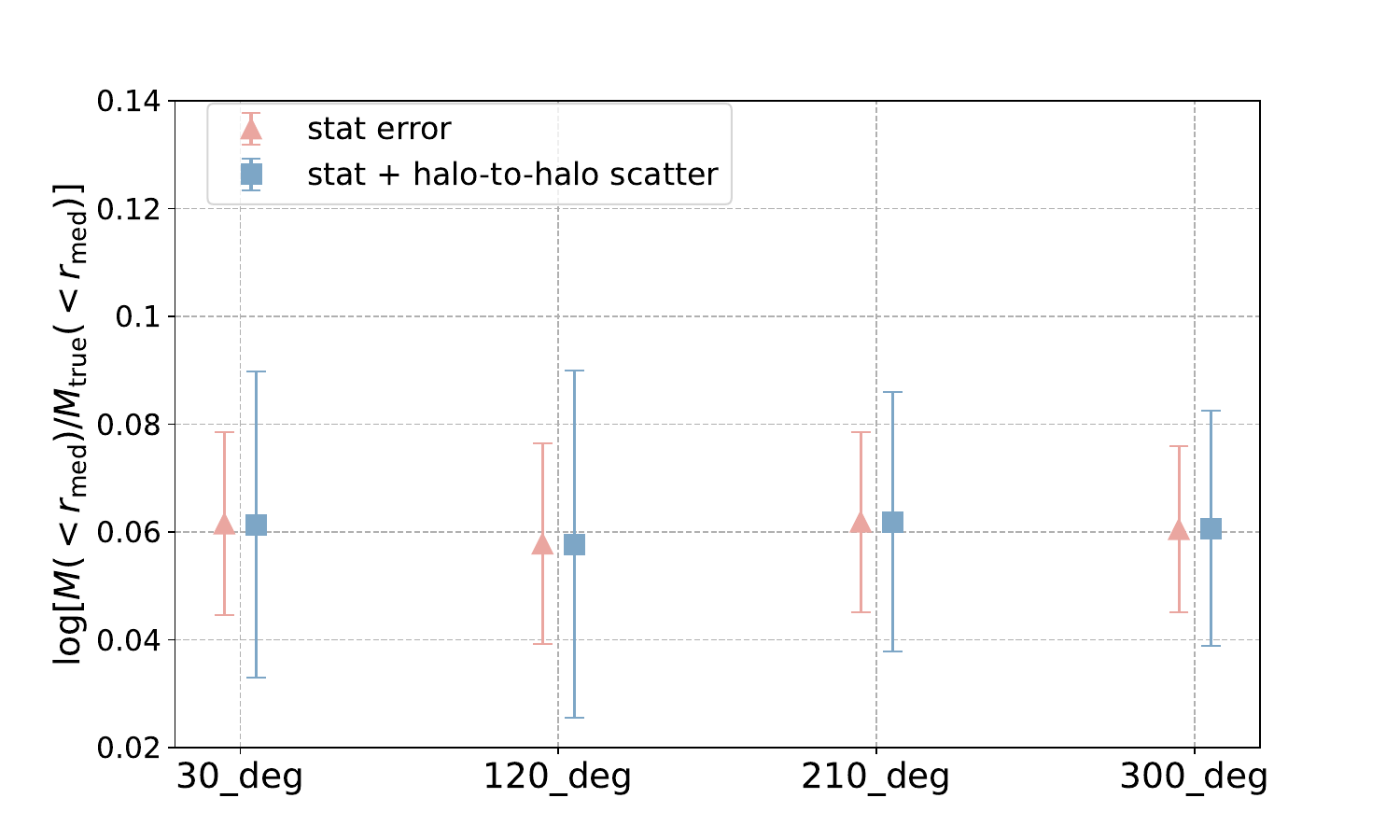}
\includegraphics[width=\columnwidth]{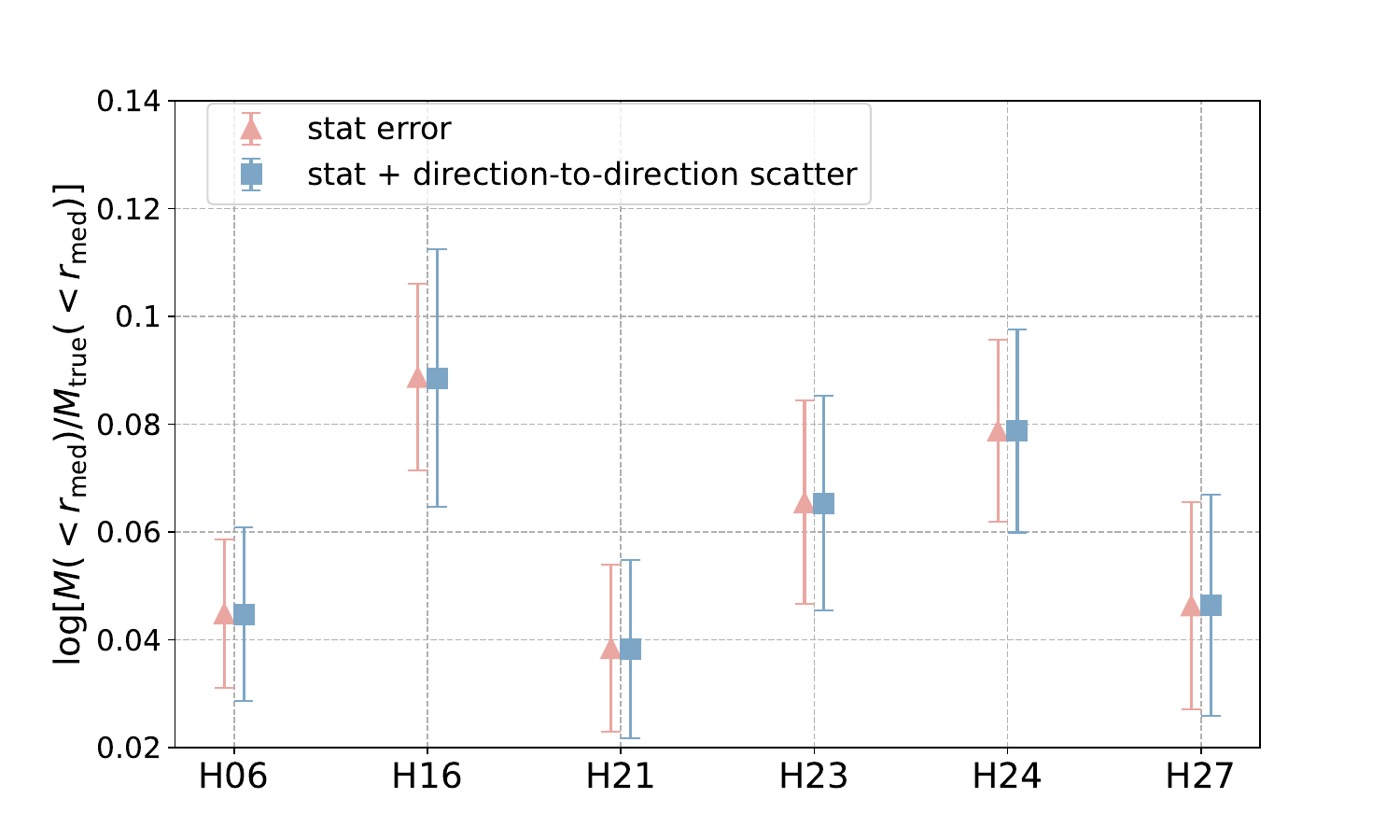}
\caption{Constraint of mass at the median tracer radius, $r_{\rm med}$. In both plots, the points represent mean values of estimated mass relative to the truth at median radius of tracers, and the size of the errorbars denote the associated uncertainties. {\bf Left:} The pink and blue errorbars show mean statistical errors and total errors of different halos with a fixed viewing direction. {\bf Right:} The pink and blue errorbars are mean statistical errors and total errors of a fixed halo with different viewing directions.} 
\label{fig:uncertainty at Mhalf}
\end{figure*}

Figure~\ref{fig:uncertainty at Mhalf} displays the difference between the estimates and truth at the median radius of tracers, together with the statistical and total uncertainty. The statistical uncertainty is obtained by averaging the formal statistical error of individual systems. The total uncertainty is derived by measuring the scatter among all halos or all directions for each case. Comparison of the two types of uncertainties, therefore, represents the additional variance introduced by the diversity among halos or different viewing directions of the same halo.

As shown in Figure~\ref{fig:uncertainty at Mhalf}, the mean bias of estimated enclosed mass at $r_{\rm med}$ relative to the truth is $\sim 0.06$ dex, the mean formal statistical error on $\log M(r_{\rm med})$ is $\sim 0.017$ dex. Both the bias and the uncertainty are quite small compared with those at the virial radius, consistent with the expectation that this radius provides the tightest  mass constraint. 

The total uncertainties taking into account the halo-to-halo scatter and the direction-to-direction scatter are both larger than the statistical uncertainty, amounting to 0.027 and 0.019 dex, respectively. Furthermore, the bias and uncertainty also vary more significantly across different halos than across different observational directions.  This aligns with our finding that the scatter in the total mass among different halos is larger than the scatter among different directions.

\subsection{Applying to DESI BHB}

\begin{figure*}
\centering
\includegraphics[width=0.96\columnwidth]{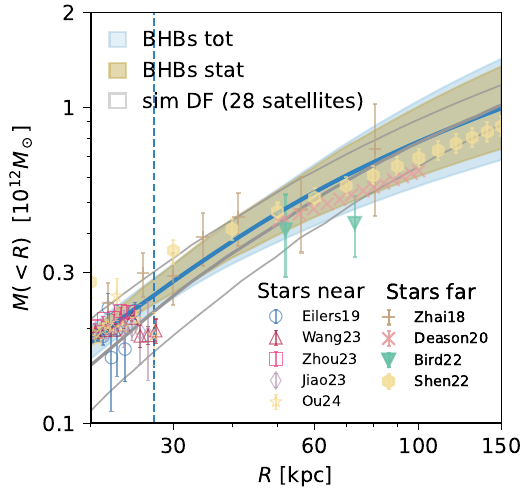}
\includegraphics[width=0.96\columnwidth]{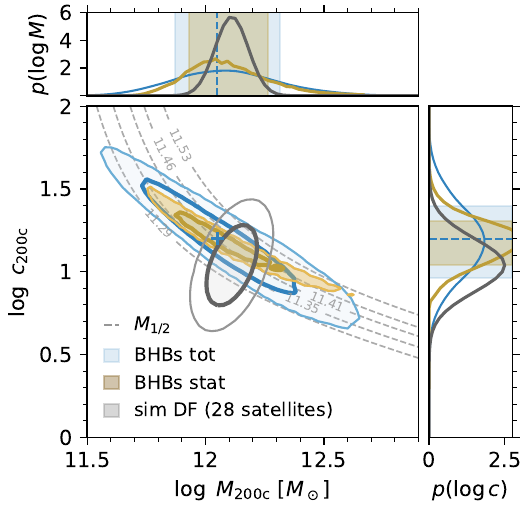}
\caption{{\bf Left:} Best constrained MW mass profiles from 
BHBs. The blue curve and shaded blue band correspond to the median estimate and 1$\sigma$ total uncertainties, with bias corrected (see Section~\ref{sec:mock-cal}) and systematic errors included. The bronze shaded band represents the formal statistical uncertainty only. The vertical dashed blue line represents the median radius of BHBs. The thick and thin gray curves show the median and 1$\sigma$ bounds from \citet{Zhaozhou2020} using the \simDF\  method with satellites as tracers. Recent studies using stars as tracers are also shown in the plot for comparison, distinguished in the legend as the stars within 30 kpc (``stars near'') and the more distant halo stars(``stars far''). {\bf Right:} Best constrained MW halo mass and concentration using BHBs. The thick and thin bronze contours represent the 1$\sigma$ and 2$\sigma$ formal statistical uncertainties, while the blue contours show the calibrated total uncertainties including systematics. Gray contours represent the result from \simDF\ for comparison. The background gray dashed lines denote constant values of enclosed masses within the median tracer radius, $M_{r_{\rm med}}$, showing the best-fit, $\pm 1\sigma$ and $\pm 2\sigma$ values including systematic errors. The numbers on top of the gray dashed lines indicate the corresponding values of $\log M_{r_{\rm med}}/M_\odot$. The upper and right subpanels display the marginalized distributions, with the dashed vertical lines showing the best-fit values in the total posterior and the shaded regions showing the $1\sigma$ uncertainties.}

\label{fig:obs corrected mainbright r0<20 BHB mass profile and m-c}
\end{figure*}

With the calibrations of biases and uncertainties derived from the mock analyses above, we apply the \empdf method to the observational DESI BHB samples. Figure~\ref{fig:obs corrected mainbright r0<20 BHB mass profile and m-c} shows the resulting enclosed mass profiles and the corresponding constraints in the halo mass-concentration plane. The results have been corrected for the estimator bias, and we show both the formal error and the total uncertainty which includes the systematic uncertainty. Tabulated mass profile constraints are also listed in Table~\ref{tab:enclosed mass}. The mass profile is derived by evaluating the posterior distribution of enclosed mass over all potential parameters at each radius. The median and $1\sigma$ interval of this distribution define the curve and its band at that radius. This provides a direct, radially local constraint that is not a simple mapping from the mass–concentration contour.
The best-fit halo parameters correspond to the maximum posterior probability within the total uncertainty contour to ensure consistency with our bias correction, which was calibrated on mock data using the mean bias of the total error constraints.

\begin{table*}[htb]
\centering
\renewcommand{\arraystretch}{1.2}
\caption{The first and second columns show the enclosed masses at different radii using BHB and RRL as tracers. The joint column on the right shows the best constraints jointly from BHBs, RRLs and the \simDF\ in \citet{Zhaozhou2020}. Formal statistical uncertainties and the total uncertainties (in bracket) are both provided. Here the last row shows the best constrained total mass within the median radius of the tracers, with $r_{\rm med}= 27.33 $ and $26.59$ kpc for BHBs and RRLs, respectively. Note our tracers extend from 20~kpc to $\sim$80 kpc.}

\begin{tabular}{lccc}
\hline
 & BHB & RRL & Joint \\
\hline
$\substack{M(<20\,\mathrm{kpc}) \\[1pt][\times10^{12}\,\mathrm{M_{\odot}}]}$ & $0.18_{-0.00}^{+0.02}(_{-0.02}^{+0.02})$ &  $0.20_{-0.01}^{+0.02}(_{-0.02}^{+0.03})$ & $0.19_{-0.00}^{+0.01}(_{-0.02}^{+0.01})$ \\[5pt]
$\substack{M(<50\,\mathrm{kpc}) \\[1pt] [\times10^{12}\,\mathrm{M_{\odot}}]}$ & $0.46_{-0.07}^{+0.08}(_{-0.09}^{+0.09})$ & $0.53_{-0.09}^{+0.12}(_{-0.10}^{+0.12})$ & $0.50_{-0.02}^{+0.02}(_{-0.03}^{+0.04})$ \\
$\substack{M(<80\,\mathrm{kpc}) \\[1pt] [\times10^{12}\,\mathrm{M_{\odot}}]}$ & $0.67_{-0.14}^{+0.17}(_{-0.17}^{+0.20})$ & $0.78_{-0.16}^{+0.22}(_{-0.20}^{+0.26})$ & $0.72_{-0.05}^{+0.05}(_{-0.05}^{+0.06})$ \\[5pt]
$\substack{M(<100\,\mathrm{kpc}) \\[1pt] [\times10^{12}\,\mathrm{M_{\odot}}]}$ & $0.78_{-0.17}^{+0.23}(_{-0.22}^{+0.27})$ & $0.91_{-0.20}^{+0.29}(_{-0.25}^{+0.34})$ & $0.85_{-0.07}^{+0.05}(_{-0.06}^{+0.07})$ \\[5pt]
$\substack{M(<150\,\mathrm{kpc}) \\[1pt] [\times10^{12}\,\mathrm{M_{\odot}}]}$ & $0.99_{-0.25}^{+0.36}(_{-0.31}^{+0.43})$ & $1.17_{-0.28}^{+0.45}(_{-0.35}^{+0.54})$ & $1.08_{-0.10}^{+0.09}(_{-0.09}^{+0.10})$ \\[5pt]
$\substack{M(<200\,\mathrm{kpc}) \\[1pt] [\times10^{12}\,\mathrm{M_{\odot}}]}$ & $1.16_{-0.32}^{+0.46}(_{-0.39}^{+0.57})$ & $1.37_{-0.35}^{+0.58}(_{-0.44}^{+0.70})$ & $1.25_{-0.12}^{+0.12}(_{-0.11}^{+0.12})$ \\[5pt]
$\substack{M(<r_{\rm med}) \\[1pt] [\times10^{12}\,\mathrm{M_{\odot}}]}$ & $0.25_{-0.02}^{+0.03}(_{-0.03}^{+0.03})$ & $0.28_{-0.02}^{+0.04}(_{-0.03}^{+0.04})$ &  \\[5pt]
\hline

\end{tabular}

\label{tab:enclosed mass}
\end{table*}

Figure~\ref{fig:obs corrected mainbright r0<20 BHB mass profile and m-c} also shows measurements from some other works for comparison. Satellite-based measurements provide a useful comparison because they probe a complementary tracer population at larger radii \citep[e.g.,][]{Patel2018,Callingham2019,Zhaozhou2020}. Here we focus on the comparison with that from \citet{Zhaozhou2020}, who modeled the phase space distribution of 28 MW satellites using the simulation-informed distribution function (\simDF) method~\citep{Li2019}.  \simDF\  assumes the satellites are in a steady-state with a distribution of orbital parameters, $p(E,L)$, following a universal function informed by cosmological distributions. The introduction of a simulation informed $p(E,L)$ helps to tighten the dynamical constraint while avoiding biases due to an unjustified DF form, as long as the simulations provide a faithful representation of the real orbital distribution. 
Overall, the best-fit mass profile from \simDF is consistent with our constraints within $1\sigma$, across the radial range covered by BHBs.  Comparing the statistical uncertainties, our uncertainties are significantly smaller in the inner regions, benefiting from the large number of stellar tracers. The errors increase at larger radii due to the decrease in the number of BHBs used as tracers. 

Several recent measurements using stars as tracers are also shown in the figure. Overall, our results are broadly consistent with these measurements within the quoted uncertainties. 
For the very inner region (20–30~kpc), our mass estimates align well with studies that model the rotation curve of disk-dominated stellar tracers\citep{Eilers2019, WangHF2023, Zhou2023, Jiao2023, Ou2024}. In the outer halo, our profile remains compatible, within the quoted uncertainties, with previous measurements based on distant halo stars~\citep{Zhai2018,Deason2021,Bird2022,Shen2022}. The result from \citet{Shen2022} shows much smaller uncertainties than ours, despite that they only modelled 168 stars using a DF method. This difference may partly reflect their adopted parametric model for the tracer distribution function, whereas our method avoids assuming a specific functional form for the tracer DF. 
We also note that the systematic uncertainties due to deviations from steady state and spherical symmetry are usually not modelled in these previous works. Considering these differences, our measurement with the bias-calibrated data-driven DF method provides a conservative yet robust constraint using halo stars.

In addition to the systematic uncertainties discussed above, the variations between individual measurements also reflect the different tracer populations, radial coverages, and modelling assumptions used in these studies. For example, the constraints from \citet{Shen2022} are based on tracer samples distributed in much larger radii, so the inferred inner mass profile depends on extrapolations that rely on the adopted potential model. \citet{Zhai2018} modeled K giants over $5$--$120~\mathrm{kpc}$ using the spherical Jeans equation, which requires assumptions about the tracer density profile and velocity anisotropy. These differences likely contribute to the method and tracer dependent scatter among current halo star mass measurements.

The right panel of Figure~\ref{fig:obs corrected mainbright r0<20 BHB mass profile and m-c} shows the posterior distribution of the virial mass and concentration parameters. The constraints show a tight degeneracy between the mass and concentration parameters. This degeneracy is equivalent to a tight constraint on the mass within the median tracer radius, which is also shown via the gray dashed contours. Parameter combinations along the degeneracy direction yield very similar $M(r_{\rm med})$, while those perpendicular to it correspond primarily to variations in $M(r_{\rm med})$. 

Compared with the \simDF\ result of \citet{Zhaozhou2020},  our constraint shows substantial overlap with the \simDF\  constraint, yet the best-fits exhibit slight offset in virial mass and concentration.  Importantly, this offset is largely aligned with the main mass-concentration degeneracy direction, so the two analyses still imply similar enclosed masses near the characteristic radius directly probed by the tracers. Along the degeneracy direction of our constraint, \simDF\ places tighter constraint than ours. This is mainly because \simDF\ also makes use of a simulation-informed $p(E,L)$ in the model, which provides complementary information to the steady-state assumption~\citep{Li2019,LiZhaozhou2024}. Besides, the two samples span different radial ranges, which also contributes to the different degeneracy directions as they constrain different $M(r_{\rm med})$. Perpendicular to our degeneracy direction, however, our constraints are much tighter than that from \simDF, due to the larger tracer sample size in our work.

The best-fit values of $\log M_\text{200c}$ and $\log c_\text{200c}$ along with their uncertainties are reported in Table~\ref{tab:obs_results}. As the posterior distributions are close to Gaussian, we also provide the parameters of a 2-dimensional Gaussian approximation to the joint posterior of $(\log M_\text{200c}, \log c_\text{200c})$, which can be easily reproduced for comparisons with other works.

\begin{table*}[ht]
\centering
\caption{Summary of the constraints from DESI BHB and RRL stars, and the joint stellar-satellite analysis. 
For each type of tracers we report (i) the best-fit values of $\log M_\text{200c}$ and $\log c_\text{200c}$, with formal statistical and total uncertainties (second and third columns from the left), and 
(ii) the parameters of a 2-dimensional  Gaussian approximation to the posterior of 
$(\log M_\text{200c}, \log c_\text{200c})$, including the means ($\mu$), standard deviations ($\sigma$), and correlation coefficient $\rho_{\rm corr}$. }
\label{tab:obs_results}
\renewcommand{\arraystretch}{1.25}
\begin{tabular}{lccccc}
\hline\hline
Tracer & 
$\log M_\text{200c}/M_{\odot}$ &
$\log c_\text{200c}$ &
Gaussian fit, stat &
Gaussian fit, tot \\
 & (stat / total) & (stat / total) & ($\mu_M\pm\sigma_M$, $\mu_c\pm\sigma_c$, $\rho_{\rm corr}$) & ($\mu_M\pm\sigma_M$, $\mu_c\pm\sigma_c$, $\rho_{\rm corr}$) \\ 
\hline
BHB &
$12.05_{-0.12}^{+0.21} \, (_{-0.18}^{+0.26})$ &
$1.20_{-0.16}^{+0.11} \, (_{-0.24}^{+0.20})$ &
$12.10\pm 0.17$, $1.18\pm 0.13$, -0.97 &
$12.09\pm 0.23$, $1.18\pm0.22$, -0.94
\\
RRL &
$12.15_{-0.14}^{+0.22} \, (_{-0.19}^{+0.26})$ &
$1.16_{-0.12}^{+0.09} \, (_{-0.18}^{+0.16})$ &
$12.20 \pm 0.19$, $1.15 \pm 0.11$, -0.98 &
$12.19 \pm 0.23$, $1.15 \pm 0.17$, -0.92 
\\
Joint &
$12.12_{-0.05}^{+0.05} \, (_{-0.04}^{+0.06})$ &
$1.16_{-0.03}^{+0.05} \, (_{-0.07}^{+0.05})$ &
$12.12\pm 0.05$, $1.17\pm 0.03$, -0.90 &
$12.14\pm 0.05$, $1.15\pm0.06$, -0.54
\\
\hline
\end{tabular}
\end{table*}

\subsection{Applying to DESI RR Lyrae}

We perform a similar mock-calibration analysis to the RRL samples. The best-fit values of $\log M_\text{200c}$ and $\log c_\text{200c}$ derived from  RRLs along with the 2D Gaussian fitting results are listed in Table~\ref{tab:obs_results}. The inferred enclosed mass profile at different radius are shown in Table~\ref{tab:enclosed mass}.

\begin{figure*}
\centering
\includegraphics[width=0.96\columnwidth]{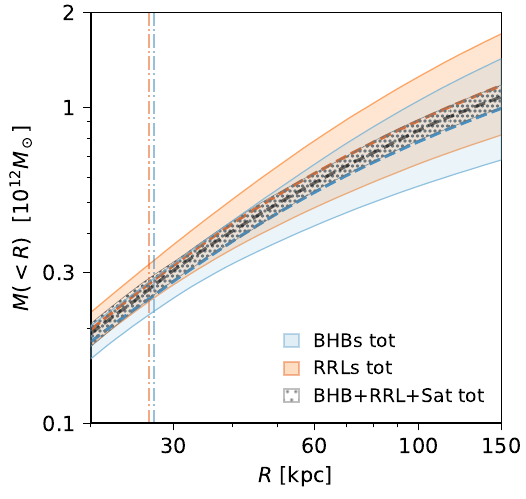}
\includegraphics[width=0.96\columnwidth]{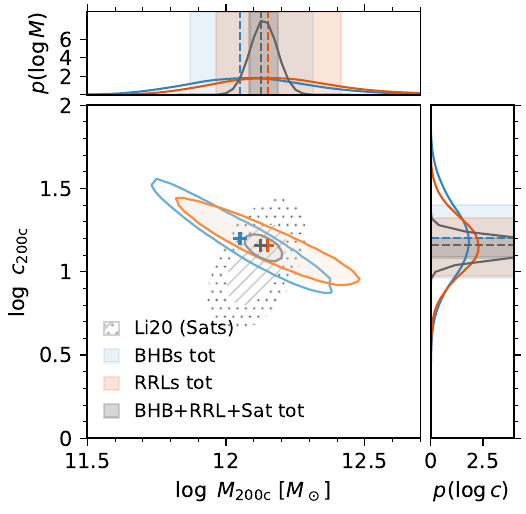}
\caption{{\bf Left:} MW mass profiles inferred from DESI main bright BHBs and RRLs. The blue and orange dashed curves with shaded bands in corresponding colors show the median estimate and 1$\sigma$ total uncertainty based on BHBs and RRLs, respectively, with biases corrected (see~Section~\ref{sec:mock-cal}) and systematic errors included. The gray line with gray dotted band represent the mean and 1$\sigma$ total uncertainty from the joint constraint of BHBs, RRLs and satellites. The vertical blue and orange dashed lines mark the median radii of BHBs and RRLs. {\bf Right:} Best constrained MW halo mass and concentration. Blue and orange crosses with shaded contours in corresponding colors denote the best constraints and 1$\sigma$ total uncertainties based on BHBs and RRLs, with systematics errors included. The gray cross and shaded contour indicate the joint constraint from halo stars plus satellites and the total 1-$\sigma$ uncertainty.  Marginalized distributions are displayed in the upper and right subpanels.}
\label{fig:obs corrected BHB and RRL mass profile and m-c}
\end{figure*}

Figure~\ref{fig:obs corrected BHB and RRL mass profile and m-c} compares the enclosed mass profiles and mass-concentration contours derived from RRLs with those obtained from the BHBs. For clarity, only the total uncertainties are shown here.  
Overall, the RRL and BHB based constraints remain generally consistent within their total uncertainties, with the RRL result favoring a modestly higher mass profile over most of the radial range shown.  This difference is also reflected in the $M-c$ plane, where the RRL constraint is shifted toward slightly higher virial mass and lower concentration relative to the BHB result. 

The enclosed mass inferred from the RRL analysis is correspondingly slightly higher than that from the BHBs, although the two remain consistent within the $1\sigma$ uncertainties. 
The two independent tracer populations produce overlapping, though not identical, constraints. This suggests that the overall inference is robust, while some variation between different tracers persists at the current precision.
Notably, the BHB and RRL contours are elongated along a very similar mass-concentration degeneracy direction, implying that the difference between them is smaller in terms of enclosed mass near the median tracer radius than it appears in the $(M,c)$ plane alone.

\subsection{Joint constraint with satellites and measurements using the \opdf}
\label{sec: compare with opdf}

To take advantage of the complementary information provided by different tracers, we combined the posterior constraints from our BHB and RRL analyses with the satellite-based constraints of \citet{Zhaozhou2020}. Specifically, we approximate the satellite posterior with a bivariate Gaussian and multiply it with the calibrated BHB and RRL posterior distributions, assuming that the three constraints are independent, to obtain the joint posterior. 
Different tracer populations probe different regions of the halo, and the distinct radial coverage and modelling assumptions lead to characteristic degeneracy directions in the parameter plane. Therefore, joint constraints help to significantly tighten the constraints. Such improvements are shown clearly in Figure~\ref{fig:obs corrected BHB and RRL mass profile and m-c}.
Combining the BHB and RRL constraints alone does not qualitatively change the degeneracy direction, since the two halo star tracers are elongated along a very similar direction in the $M-c$ plane, although it does improve the overall constraint. Once the satellite constraint which has a more complementary orientation in parameter space is further included, the posterior shrinks substantially and the degeneracy is greatly reduced. 
The resulting joint mass profile becomes more tightly constrained particularly at larger radii, where the satellites contribute most strongly. The joint constraint is therefore driven both by the consistency of the two stellar tracers and by the complementary parameter space orientation of the satellite-based result.
As both methods involve minimal assumptions that are well tested and calibrated in simulations, our combined analysis provides one of the tightest and most robust measurements of the mass profile of the MW.

Finally, we repeat the analysis on the same BHB and RRL samples using the precursor of \empdf, the \opdf method, which can help to cross check with the constraints obtained above. In Appendix~\ref{sec:opdf}, we show that the results from \opdf are broadly in agreement with those from the \empdf, but with generally larger uncertainty, consistent with the fact that \empdf uses a smoothed empirical DF and a full phase-space likelihood.

\section{Conclusions}

In this work, we constrain the Milky Way’s mass profile using the data-driven dynamical modeling method \empdf, applied to a large sample of halo stars from the DESI MWS. Specifically, we utilize BHB and RR Lyrae stars as dynamical tracers owing to their accurate distance estimates and widespread presence in the outer stellar halo. Our analysis begins with tests on the DR2 AuriDESI mock catalogues presented in \citet{Riley:2026auridesi}, to calibrate estimator uncertainties and biases and quantify systematics across different halos and viewing angles. The calibrated estimator is then applied to the real DESI observations. This analysis forms part of the broader DESI DR2 Milky Way mass Key Project, alongside complementary mass profile measurements from other dynamical methods \citep{Medina:2026gme,Valluri:2026nimble} and the overall synthesis presented in \citet{DESI:2026mwmass}. The key findings and conclusions are summarized as follows:

\begin{itemize}
\item In the absence of observational errors, \empdf can recover the true mass profile in an ensemble unbiased manner. However, when realistic observational errors are incorporated, an overall bias emerges, particularly due to the increasing uncertainties in the tangential velocity measurements at larger distances. Modeling on mock data helps to calibrate the results from real data, including the correction of both systematic bias and  scatter.

\item 
By applying our method to mock samples, we find that the total scatter of the best-fit parameters is much larger than the average statistical error, indicating the presence of systematic errors. This primarily originates from the halo-to-halo scatter. 
Nevertheless, the constraint on the enclosed mass at the median tracer radius remains remarkably tight even after including systematics, confirming that this characteristic radius provides the strongest dynamical constraint.

\item Applying the calibration derived from the mocks to the observational BHB data, we obtain $\log M_\text{200c}/M_{\odot}=12.05_{-0.12}^{+0.21} \left(_{-0.18}^{+0.26}\right)$ and $\log c_\text{200c}=1.20_{-0.16}^{+0.11}\left(_{-0.24}^{+0.20}\right)$. The corresponding enclosed mass at 100~kpc is $M(<100\,\mathrm{kpc}) = 0.78_{-0.17}^{+0.23}(_{-0.22}^{+0.27}) \times 10^{12}\,M_{\odot}$. The BHB-based profile is broadly consistent with previous constraints from halo stars within the quoted uncertainties.

\item Applying similar calibration to observational RR Lyrae data, the best constrained mass and concentration are $\log M_\text{200c}/M_{\odot}=12.15_{-0.14}^{+0.22}\left(_{-0.19}^{+0.26}\right)$ and $\log c_\text{200c}=1.16_{-0.12}^{+0.09}\left(_{-0.18}^{+0.16}\right)$. The RRL and BHB results are closely consistent within their uncertainties, providing mutually consistent mass profile constraints from two independent halo tracer populations.

\item  The constraints from our analysis are highly complementary to previous constraints using satellite dynamics. The mass profile constraints from our halo star samples are tightest in the inner halo at $\sim 30$ kpc, with a total uncertainty of $\sim 0.03$ dex. This corresponds to a tight degeneracy in the derived mass and concentration parameters, perpendicular to the previous constraint from \citet{Zhaozhou2020} using satellite dynamics. By combining the constraints from the stars with those from the satellites, we obtain a joint posterior that provides one of the tightest yet well-calibrated constraints to date on the MW halo mass, with the best-fit $\log M_\text{200c}/M_{\odot}=12.12_{-0.05}^{+0.05}\left(_{-0.04}^{+0.06}\right)$ and $\log c_\text{200c}=1.16_{-0.03}^{+0.05}\left(_{-0.07}^{+0.05}\right)$.

\end{itemize}
Our results also provide an external perspective on recent claims of a declining Milky Way rotation curve. Several measurements derived from rotation curve have suggested that the circular velocity declines beyond $\sim 15-20$ kpc, with some studies finding a particularly steep drop that may even approach a near-Keplerian behaviour \citep{WangHF2023,Jiao2023,Ou2024}. Such a rapidly declining outer rotation curve would point to an unusually low Milky Way halo mass, in tension with many independent constraints based on halo stars and satellites. However, recent work has shown that Jeans based inferences of the rotation curve can be sensitive to departures from axisymmetry and equilibrium, and to assumptions about the tracer density profile, all of which can bias the inferred decline at large radii \citep{Koop2024,Ou2025}. In this context, our constraints, which probe the MW mass profile on outer halo scales using halo stars, do not favor such a low Milky Way mass implied by the most extreme declining rotation curve interpretations. The sharpest low mass inferences are likely sensitive to modelling assumptions and systematics. Our results are instead more consistent with the broader set of other dynamical measurements. 

We also applied the \opdf method to the same observational samples for validation and comparison. The inferred mass profiles derived from the \opdf are generally consistent with those from the \empdf, demonstrating that our inference is robust. This consistency further supports the stability and reliability of our constraints on the MW mass. 

A natural extension to this work would involve implementing a framework to inherently account for observational error, such as the tentative approach outlined in Appendix~C in \cite{LiZhaozhou2024}. This would eliminate the need for external calibration from specific mock data. However, we note that simulation calibrations will remain crucial for estimating the systematic uncertainties due to deviation from steady state and spherical symmetry. Because of this, mock building efforts which can produce a much larger number of independent mock observations than AuriDESI will prove helpful for future analysis. We have already shown that combining halo stars with satellites yields a much tighter constraint, and additional tracers such as globular clusters could further strengthen the joint analysis. More broadly, the complementarity among different tracers and dynamical methods suggests that a more systematic combination of independent Milky Way mass constraints will be an important direction for future work. Another useful extension would be to adopt more general Galactic potential models. In particular, the centrally concentrated stellar and gaseous mass could be modeled separately from the extended halo, for example through a central monopole term or a more detailed multi-component potential. Such an extension would allow us to assess more directly how the treatment of the central baryonic mass affects the inferred halo scale radius and concentration. Together with analyses based on a larger suite of MW-like simulations, these developments would help better disentangle and quantify the various sources of systematic uncertainty.

In conclusion, this study demonstrates the power of combining the self data-driven method \empdf, which can fully utilize 6D phase space information of dynamical tracers, with the abundant data of halo stars provided by DESI MWS. With careful calibration based on mock analyses, \empdf delivers precise mass estimates that incorporate systematic uncertainties. As the DESI survey progresses, expanding the sample size and refining observational data, a tighter constraint of MW mass is anticipated in the future. 

\section*{Data availability}

The machine-readable data underlying the figures have been deposited on Zenodo and will be publicly available upon publication (\url{https://zenodo.org/records/22644507}).

\begin{acknowledgments}

This work is supported by NSFC (12595312, 12573022, 12022307), the National Key R \& D Program of China (2023YFA1605600, 2023YFA1605601, 2023YFA1607800, 2023YFA1607801), 111 project (No.\ B20019), the science
research grants from the China Manned Space Project (No.CMS-CSST-2025-A04), and the Office of Science and Technology, Shanghai Municipal Government (grant Nos. 24DX1400100,ZJ2023-ZD-001). We thank the sponsorship from Yangyang Development Fund. Z.Z.L. acknowledges the Fundamental Research Funds for the Central Universities (KG202502), and the Marie Sk{\l}odowska-Curie Actions Fellowship (101109759, CuspCore) under Horizon Europe. CAP acknowledges financial support from the
Spanish Ministry of Science, Innovation and Universities (MICIU) projects
PID2020-117493GB-I00, PID2023-149982NB-I00 and PID2023-146453NBI00.
AHR was supported by a fellowship funded by the Wenner-Gren Foundations (Wenner-Gren Stiftelserna), a Research Fellowship from the Royal Commission for the Exhibition of 1851, and by the Science \& Technology Facilities Council (STFC) through grant ST/T000244/1. GEM acknowledges support from the Dutch Research Council (NWO) through the Veni grant VI.Veni.252.355. GEM and TSL acknowledge financial support from Natural Sciences and Engineering Research Council of Canada (NSERC) through grant RGPIN-2022-04794.


The computations of this work are carried on the Gravity supercomputer at the Department of  and the National Energy Research Scientific Computing Center (NERSC).

This material is based upon work supported by the U.S. Department of Energy (DOE), Office of Science, Office of High-Energy Physics, under Contract No. DE–AC02–05CH11231, and by the National Energy Research Scientific Computing Center, a DOE Office of Science User Facility under the same contract. Additional support for DESI was provided by the U.S. National Science Foundation (NSF), Division of Astronomical Sciences under Contract No. AST-0950945 to the NSF’s National Optical-Infrared Astronomy Research Laboratory; the Science and Technology Facilities Council of the United Kingdom; the Gordon and Betty Moore Foundation; the Heising-Simons Foundation; the French Alternative Energies and Atomic Energy Commission (CEA); the Secretariat of Science, Humanities, Technology and Innovation
(SECIHTI) of Mexico; the Ministry of Science, Innovation and Universities of Spain (MICIU/AEI/10.13039/501100011033), and by the DESI Member Institutions: \url{https://www.desi.lbl.gov/collaborating-institutions}.Any opinions, findings, and conclusions or recommendations expressed in this material are those of the author(s) and do not necessarily reflect the views of the U. S. National Science Foundation, the U. S. Department of Energy, or any of the listed funding agencies.

The authors are honored to be permitted to conduct scientific research on I'oligam Du'ag (Kitt Peak), a mountain with particular significance to the Tohono O’odham Nation.

For the purpose of open access, the author has applied a Creative
Commons Attribution (CC BY) licence to any Author Accepted
Manuscript version arising from this submission.

ChatGPT (OpenAI; GPT-5.5) was used during manuscript preparation to assist with language editing and limited code debugging. All scientific analyses, interpretations, references, and final manuscript content were reviewed and verified by the authors.

\software{ \textsc{astropy} \citep{Astropy:2013, Astropy:2018, Astropy:2022}, 
\textsc{jupyter} \citep{jupyter},
\textsc{numpy} \citep{numpy},
\textsc{pandas} \citep{pandas},
\textsc{matplotlib} \citep{matplotlib},
\textsc{h5py} \citep{h5py},
\textsc{scipy} \citep{Scipy},
ChatGPT (GPT-5.5)
\citep{chatgpt2026}}

\end{acknowledgments}
\appendix
\section{Halo-to-halo and directional systematic uncertainties}
\label{app:sys_component}

\begin{figure*}
\centering
\includegraphics[width=1.6\columnwidth]{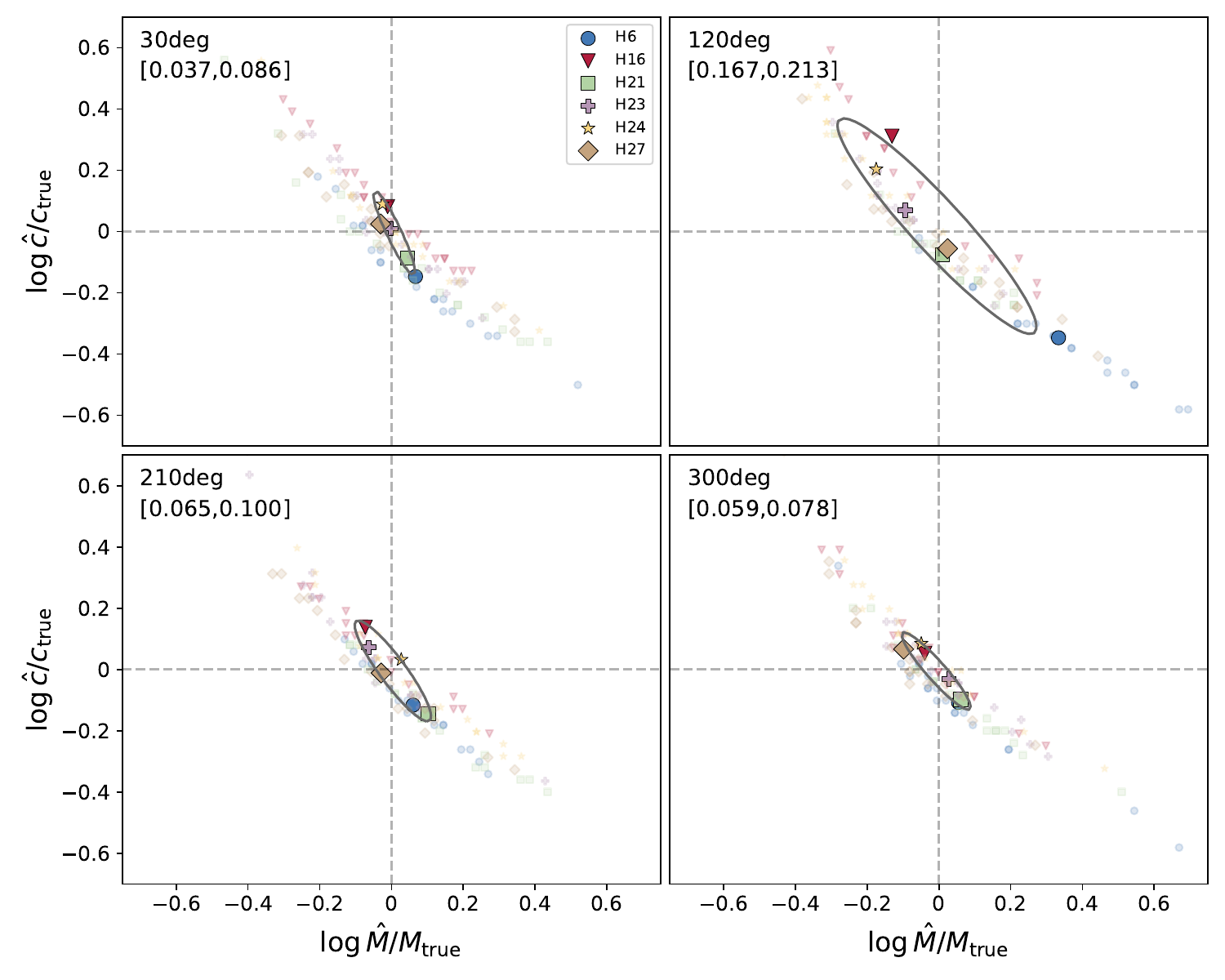}
\caption{Bias-corrected residuals in the best-fit halo mass ($M$) and concentration ($c$) for the error-added BHB mock samples. Each panel fixes one viewing direction and compares the six mock halos. Faded symbols show the individual error realizations, while the larger symbols show the mean over the 20 realizations for each halo. The gray ellipse is computed only from the larger symbols and therefore represents the halo-to-halo scatter at fixed viewing direction. The numbers in brackets give the root-mean-square error of mass and concentration for the realization-averaged estimates in each panel.}
\label{fig: bestfit halo-halo sys}
\end{figure*}

\begin{figure*}
\centering
\includegraphics[width=1.92\columnwidth]{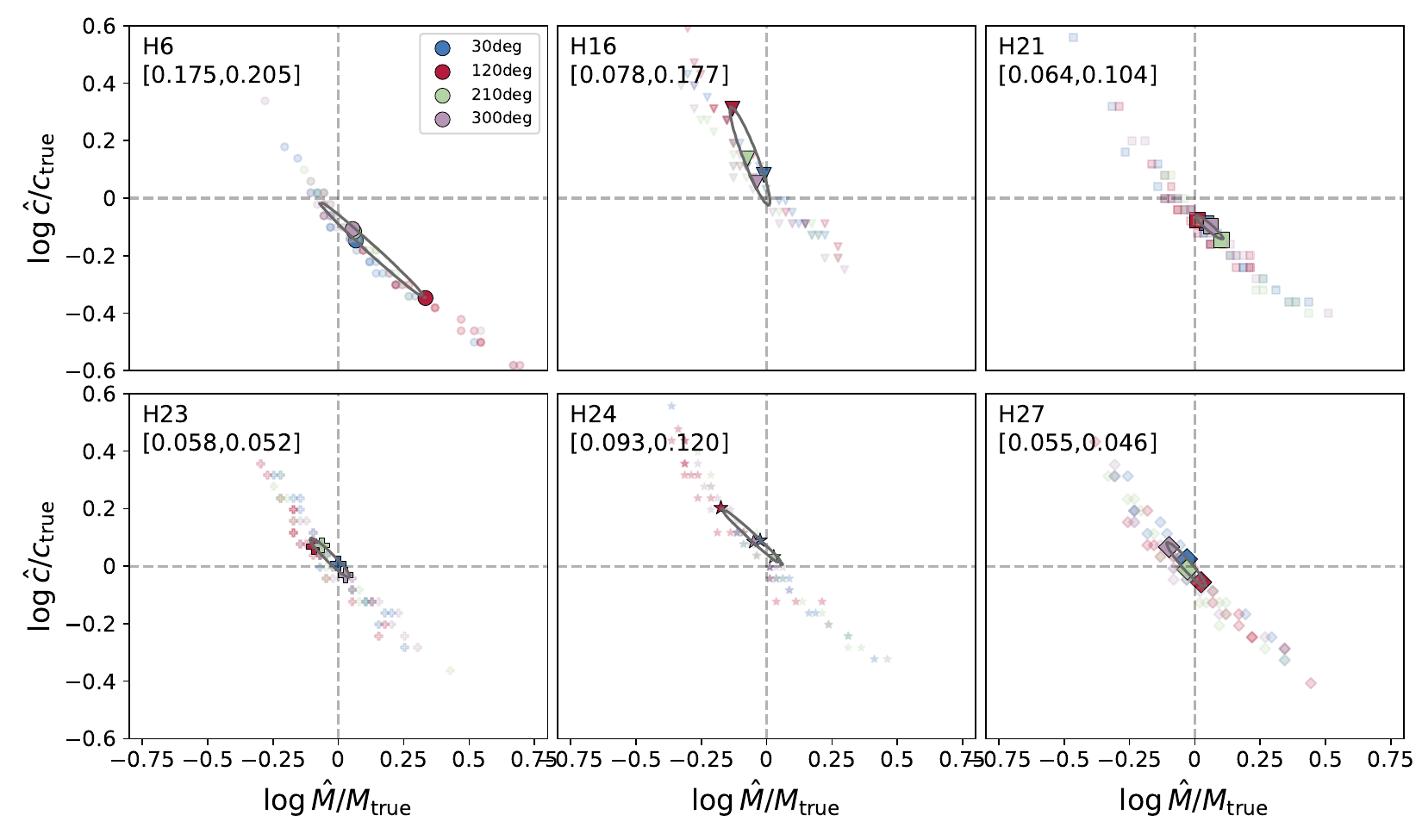}
\caption{Similar to Figure~\ref{fig: bestfit halo-halo sys}, but each panel fixes one halo and compares the four viewing directions. Faded symbols show the individual error realizations, while the larger symbols show the mean over the 20 realizations for each viewing direction. The gray ellipse is computed only from the larger symbols and therefore represents the direction-to-direction scatter at fixed halo. The root-mean-square error of mass and concentration for the larger symbols are provided in the bracket of the upper left corner in each panel.}
\label{fig: bestfit deg-deg sys}
\end{figure*}

We further decompose the systematic uncertainty using the error-added BHB mock samples. For each fixed halo and viewing direction, we first average the recovered parameters over the 20 error realizations. This removes the stochastic scatter associated with individual error realizations and allows us to focus on the variations between different halos and between different viewing directions. 
In Figure~\ref{fig: bestfit halo-halo sys}, we obtain the estimates for different halos while keeping the viewing direction fixed. 
The estimates shown in this plot have been corrected for the net bias, allowing us to focus on the scatter. For fixed viewing direction, the averaged estimate of different halos remain broadly distributed around the origin, suggesting the presence of halo-to-halo scatter. Similarly, for the same halo observed from different directions, the scatter of the averaged estimate represents the direction-to-direction scatter, as shown in Figure~\ref{fig: bestfit deg-deg sys}. Notably, for each fixed viewing direction, the ensemble of halos does not show a strong coherent offset in any particular viewing direction. While for a fixed halo, the estimates from different viewing directions often remain in a similar region of the $(\log M,\log c)$ plane for a given halo, rather than being symmetrically distributed around the origin. This suggests that each halo can have its own baseline offset. Changing the viewing direction moves the estimates around this halo-dependent baseline but does not fully eliminate it. 

Quantitatively, the mean halo-to-halo scatter measured from the realization-averaged centroids is $[\sigma_{\log M},\sigma_{\log c}]_{\rm halo}$ = [0.090,0.130], where the average is taken over the four fixed viewing directions. The corresponding direction-to-direction scatter is $[\sigma_{\log M},\sigma_{\log c}]_{\rm dir}$ = [0.070,0.072], 
 where the average is taken over the six fixed halos. The halo-to-halo scatter is therefore larger in both parameters, This indicates that variations among halos constitute the dominant source of systematic uncertainty, although the contribution from viewing-direction variation remains non-negligible.

\section{Results from \opdf}
\label{sec:opdf}

\begin{figure*}
\centering

\includegraphics[width=0.96\columnwidth]{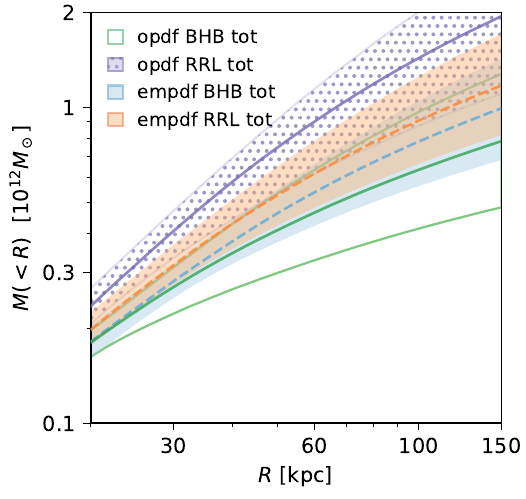}
\includegraphics[width=0.96\columnwidth]{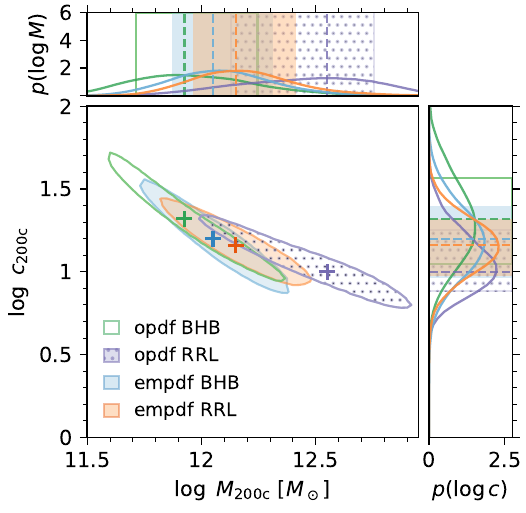}
\caption{Comparison between the results obtained from \empdf and \opdf. {\bf Left:} MW mass profiles inferred from both methods. The green curve with light green boundary and purple curve with dotted purple band represent the \opdf results for BHBs and RRLs, respectively. Blue and orange dashed curves with shaded band in corresponding colors are based on \empdf and are for BHBs and RRLs. Only the total uncertainties including systematic errors are shown here.
{\bf Right:} Mass and concentration inferred by \empdf and \opdf (see the legend). The filled contours are $1 \sigma$ total uncertainties. The upper and right subpanels show the marginalized distributions.}
\label{fig:opdf obs}
\end{figure*}

We calibrate the \opdf results independently using the AuriDESI mocks, following the same general procedure as for \empdf. The final results are compared in Figure~\ref{fig:opdf obs}. Overall, the mass profiles inferred from \opdf are broadly consistent with those from \empdf, although the details of the shifts differ between tracers. The \opdf constraints are generally broader, while the two tracers remain statistically consistent within the total uncertainties for both estimators. The corresponding constraints in the mass-concentration plane are also consistent within the total uncertainty contours. These results support the robustness of the inferred MW mass profile against the choice of dynamical estimator.

\bibliography{ref}{}
\bibliographystyle{AAStemp_v7/aasjournalv7.1}
\end{document}